\documentclass[10pt,paper]{JHEP}
\usepackage{graphicx}% Include figure files
\usepackage{dcolumn}% Align table columns on decimal point
\usepackage{bm}% bold math
\usepackage{amsfonts}
\usepackage{amsthm}
\usepackage{amsmath}
\usepackage{amssymb}
\usepackage{cite}
\usepackage{epsfig}
\newcommand{\bsl}{\begin{split}}
\newcommand{\esl}{\end{split}}

\newcommand{\cd}{{D}}
\newcommand{\bea}{\begin{eqnarray}}
\newcommand{\eea}{\end{eqnarray}}
\newcommand{\be}{\begin{equation}}
\newcommand{\ee}{\end{equation}}
\newcommand{\half}{\frac{1}{2}}
\newcommand{\bi}{\begin{itemize}}
\newcommand{\ei}{\end{itemize}}

\newcommand{\rad}{ \left( \frac{\pi^2}{30}\right) g_\ast T^4 }

\newcommand{\radw}{ \left( \frac{\pi^2}{30}\right) g_\ast }

\newcommand{\ba}{\begin{eqnarray}}
\newcommand{\ea}{\end{eqnarray}}
\newcommand{\notE}{E\kern-0.6em\hbox{/}\kern0.05em}
\newcommand{\notEt}{E_{T}\kern-1.21em\hbox{/}\kern0.45em}

\def\bi{\begin{itemize}}
\def\ei{\end{itemize}}

\def\P{P_{\rm eff}}

\preprint{UCB-PTH-08/06 \\ MCTP-08-10}

\title{Non-thermal Dark Matter and the Moduli Problem\\
in String Frameworks}
\author{Bobby S. Acharya\footnote{\tt{bacharyaATcern.ch}}\\
Abdus Salam International Center for Theoretical Physics \\
Strada Costiera 11,34014 Trieste. ITALY and \\
INFN, Sezione di Trieste and \\
Michigan Center for Theoretical Physics\\
University of Michigan, \\ Ann Arbor, Michigan 48109, USA}

\author{Piyush Kumar\footnote{\tt{kpiyush@berkeley.edu}}\\
Department of Physics\\University of California, Berkeley, CA
94720, USA and \\Theoretical Physics Group\\ Lawrence Berkeley
National Laboratory, Berkeley, CA 94720, USA}

\author{Konstantin Bobkov\footnote{\tt{bobkov@umich.edu}},
Gordon Kane\footnote{\tt{gkane@umich.edu}},
Jing Shao\footnote{\tt{jingshao@umich.edu}}, and Scott Watson\footnote{\tt{watsongs@umich.edu}}
\\Michigan Center for Theoretical Physics\\
University of Michigan, \\ Ann Arbor, Michigan 48109, USA}

\abstract{We address the cosmological moduli/gravitino problems
and the issue of too little thermal but excessive non-thermal dark
matter from the decays of moduli. The main examples we study are
the $G_2$-MSSM models arising from $M$ theory compactifications,
which allow for a precise calculation of moduli decay rates and
widths. We find that the late decaying moduli satisfy both BBN
constraints and avoid the gravitino problem. The non-thermal
production of Wino LSPs, which is a prediction of $G_2$-MSSM
models, gives a relic density of about the right order of
magnitude. }

\begin{document}

\section{Introduction}

The existence of Dark Matter seems to require physics beyond the
Standard Model. If this physics arises from a string/$M$ theory
vacuum, one is faced with various problems associated with the
moduli fields, which are
gauge-singlet scalar fields that arise when compactifying
string/$M$ theory to four dimensions. 
In particular, moduli
fields can give rise to disastrous cosmological effects.

For example, the moduli have to be stabilized, or made massive, in
accord with cosmological observations. Even if these moduli are
made massive, there could be a large amount of energy stored in
them leading to the formation of scalar condensates. In most
cases, this condensate will scale like ordinary matter and will
quickly come to dominate the energy density. The moduli are
unstable to decays to photons, and when this occurs, the resulting
entropy can often spoil the successes of big-bang nucleosynthesis
(BBN). This is the cosmological moduli problem
\cite{Coughlan:1983ci,Ellis:1986zt,hep-ph/9308325,hep-ph/9308292,Nakamura:2007wr}.
In supersymmetric extensions of the standard model, the
overproduction of gravitinos can cause similar problems and have been a source of much investigation
\cite{Khlopov:1984pf,Dine:2006ii,Endo:2006qk,Kawasaki:2006hm,Nakamura:2006uc,Asaka:2006bv,Endo:2006zj,Rychkov:2007uq,hep-ph/0602061,hep-ph/0602081,hep-ph/0605091,arxiv:0706.0986,hep-ph/0701042,astro-ph/0505395}. 

In addition, the ``standard" picture in which Dark Matter (DM)
particles are produced during a phase of thermal equilibrium can
be significantly altered in the presence of moduli.  The moduli,
which scale like non-relativistic matter, typically dominate the
energy density of the Universe making it matter dominated.
Therefore, the dominant mechanism for production of DM particles
is non-thermal production via the direct decay of moduli\footnote{For other phenomenologically 
based approaches to non-thermal dark matter and the related issue of baryon asymmetry in the 
presence of scalar decay see \cite{Nagai:2007ud,Kawasaki:2007yy,hep-ph/0606075}.}. However,
this can lead to further problems since it is easy to produce too
much dark matter compared with what we observe today. 

In recent years there has been considerable progress in our
understanding of moduli dynamics and their potential in different
frameworks which arise in various limits of string/$M$ theory. The
most popular examples include the KKLT and Large Volume frameworks
in Type IIB string theory
\cite{hep-th/0301240,hep-th/0502058,hep-th/0505076}, where all
moduli are stabilized by a combination of fluxes and quantum
corrections. These frameworks are also attractive in the sense
that they provide a mechanism for supersymmetry breaking at low
scales ($\sim \mathrm{TeV}$), thus accommodating the hierarchy
between the Electroweak and Planck scales (see
\cite{Douglas:2006es,Hertzberg:2007ke,McAllister:2007bg} for
reviews). Since one can concretely study the couplings between
moduli and matter fields, we have an opportunity to address many
issues in particle physics and cosmology from an underlying
microscopic viewpoint. The cosmological moduli/gravitino problems
and adequate generation of dark matter within the Type IIB
frameworks has met with some mixed success in a recent paper
\cite{arXiv:0705.3460}.

In this paper we will study a different framework, in which
we will also address the Dark Matter and moduli/gravitino problems.
This is the
low energy limit of $M$ theory vacua in which the extra dimensions form
a manifold of $G_2$ holonomy. Although the study of such vacua has proven to be
technically challenging, much progress has been made towards
understanding the effective four dimensional physics emerging from
them \cite{hep-th/0203061,hep-th/0109152,hep-th/0108165,atiyah}.
This includes many phenomenological implications
of these vacua, in particular relating to issues such
as constructing a realistic visible sector with chiral matter and
non-abelian gauge bosons, supersymmetry breaking, moduli
stabilization in a dS vacuum as well as explaining the Hierarchy
between the Electroweak and Planck scales,
as exemplified in
a number of works
\cite{moduli,hep-ph/0201018,hep-th/0211269,hep-th/0606262,hep-th/0701034,Acharya:2008zi}.

We will show that the moduli, gravitino and dark matter problems
are all naturally solved within this framework. Because of the
presence of moduli, the Universe is matter-dominated from the end
of inflation to the beginning of BBN. The LSPs are mostly produced
non-thermally via moduli decays. The final result for the relic
density only depends on the masses and couplings of the lightest
of the moduli (which decay last) and the mass of the LSP. This is
related to the fact that the LSP is a Wino in the $G_2$-MSSM and
that there is a fairly model independent critical LSP density at
freeze out. For natural/reasonable choices of microscopic
parameters defining the $G_2$ framework, one finds that it is
possible to obtain a relic density of the right order of magnitude
(up to factors of ${\cal O}(1)$). With a more sophisticated
understanding of the microscopic theory, one might obtain a more
precise result. The qualitative features which are crucial in
solving the above problems may also be present in other realistic
string/$M$ theory frameworks.

Moduli which decay into Wino LSPs have been considered previously
in the context of Anomaly Mediated Supersymmetry Breaking Models
(AMSB) by Moroi and Randall \cite{Moroi:1999zb}. The moduli and
gaugino masses they consider are qualitatively similar to those of
the $G_2$-MSSM. There are some important differences however. In
particular, the MSSM scalar masses in the \cite{Moroi:1999zb} are
much lower than the $G_2$-MSSM, leading to much fewer LSPs
produced per modulus decay compared to the $G_2$ models.
Furthermore, unlike in AMSB, in the $G_2$ case one is able to
calculate all the moduli masses and couplings explicitly which
leads to a more detailed understanding. In essence, though, many
of the important ideas in our work are already present in
\cite{Moroi:1999zb}. The $G_2$-MSSM models can be thought of as a
concrete microscopic realization of the relevant qualitative
features of the AMSB models.

Interestingly, our actual result for the relic density (equation
\ref{LSPlight}) is a few times larger than the WMAP value if we
use central values for the microscopic constants, which should
probably be regarded as a success.  It is also worth remarking
that, contrary to common views, it is not at all possible to get
any value one wants -- we can barely accommodate the actual
observed value in the $G_2$ framework.

The paper is organized as follows. In Section \ref{asection} we
briefly summarize early universe cosmology in the presence of
moduli, and address many of the issues associated with their
stabilization and decay. In Section \ref{overview} we give a non-technical
overview of the main results. This is largely because much of this
paper involves technical calculations. In section \ref{reviewg2} we present a
brief review of the $G_2$-MSSM, a model which arises after
considering moduli stabilization within the framework of $M$
theory compactifications. A basic discussion of decay rates and
branching ratios for the moduli and gravitinos in this model
follows, with a detailed calculation left for Appendix
\ref{appendixa}. Then in section \ref{sectionresults}, we consider
again the cosmology of moduli presented in section
{\ref{asection}) for the case of the $G_2$-MSSM. In section
\ref{darkmattersection}, after a review of dark matter production
in both the thermal and non-thermal cases, we consider the dark
matter abundance arising from the non-thermal decay of the
$G_2$-MSSM moduli.  This section is a more technical overview of
the salient features of dark matter production, leaving an even
more detailed treatment for Appendix \ref{appendixb}. In this
section we present our main result, which is that the $G_2$-MSSM
naturally predicts a relic density of Wino-like neutralinos of
about the right magnitude in agreement with observation. This is
followed by a detailed discussion of the results obtained and how
it depends on the qualitative (and quantitative) features of the
underlying physics. We then conclude with considerations for the
future.

\section{Early Universe Cosmology in the Presence of Moduli \label{asection}}

Before considering the particular case of moduli in the
$G_2$-MSSM, we first briefly review the early universe evolution
of moduli and the associated cosmological issues that can result.
This section will also serve to set our conventions.

Currently, the only convincing model leading to a smooth, large,
and nearly isotropic Universe as well as providing a mechanism for
generating density perturbations for structure formation is
cosmological inflation. At present we have very little
understanding of how the ``inflationary era" might arise within
the $M$ theory framework. In what follows,therefore, we will
assume that adequate inflation and (p)reheating have taken place
and focus on the post-reheating epoch. We will also conservatively
take the inflationary reheat temperature to be near the
unification scale $10^{14}-10^{15}$ GeV, so that possibilities for
high-scale baryogenesis exist. We will comment more on this issue
at the end.

During inflation, the moduli fields are generically displaced from
their minima by an amount of ${\cal O}(m_p)$\cite{hep-ph/9507453}.
This can be seen by looking at the following generic potential
experienced by the moduli: \ba \label{modulipotential} V(\psi)
\sim \half m_{soft}^2 (\psi-\psi_0)^2 -H_{inf}^2 (\psi-\psi_0)^2 +
\frac{1}{m_p^{2n}} (\psi-\psi_0)^{4+2n} \ea where $\psi_0$ is the
true vacuum-expectation-value (vev) of the field, i.e. in the
present Universe. Only the first term in (\ref{modulipotential})
comes from zero-temperature supersymmetry breaking, the other two
highlight the importance of high-scale corrections and the
mass-squared parameter ($\sim - H_{inf}^2$) which results from the
finite energy density associated with cosmological inflation \cite{hep-ph/9507453}. As
argued earlier, the potential (\ref{modulipotential}) is dominated
by the last two terms during inflation since $H_{inf} \gg m_{soft}
\sim m_{3/2}$. Thus, a minimum of the potential will occur near:
\be \langle \psi \rangle_{inf} \sim \psi_0+m_p \left(
\frac{H_{inf}}{m_p} \right)^{1/(n+1)} \;\;\;\;\;\;\;\; H \gg
m_{soft}. \ee Here, for simplicity, we have implicitly assumed
that the induced mass-squared parameter for $\psi$ during
inflation is \emph{negative} and of $\mathcal{O}(H_{inf}^2)$. This
is possible for a non-minimal coupling between the inflationary
fields and the moduli, a generic possibility within string theory.
A large displacement of moduli fields is also possible when the
induced mass-squared parameter during inflation is positive, but
much smaller than $|H_{inf}^2|$. In this case, large dS
fluctuations can drive the moduli fields to large values during
inflation. Therefore, independent of details, the assumption we
make is that gauge singlet scalar fields like moduli (and meson
fields in the $G_2$-MSSM) will be displaced from their present
minimum by large values.

After the end of inflation and subsequent cosmological evolution,
when $H \lesssim m_{3/2}$, the soft mass term in the potential
will dominate and we have: \be \langle \psi \rangle_{present} \sim
\psi_0 \;\;\;\;\;\;\;\; \;\;\;\;   H \lesssim m_{soft}. \ee
$\psi_0$ is also typically of order $m_p$. In Section
\ref{reviewg2}, we will present the soft masses and decay rates
for the moduli arising from soft SUSY breaking in the $G_2$-MSSM
low-energy effective theory relevant in the present Universe.
Thus, we see that by considering moduli in the early universe with
high-scale inflation, it is a rather generic consequence to expect
moduli to be displaced from their low-energy (present) minimum by
an amount: \bea \label{initial} |\Delta \psi| \equiv |\langle \psi
\rangle_{inf}-\langle \psi \rangle_{present}| \approx
m_p\left(\frac{H_{inf}}{m_p}\right)^{1/(n+1)}\lesssim m_p \eea

\subsection{Addressing the ``Overshoot Problem"}

The evolution of moduli after the end of inflation is governed by
the following equation: \be \label{seom} \ddot{\psi} + (3
H+\Gamma_\psi) \dot{\psi} + \frac{\partial V}{\partial \psi}=0.
\ee where the modulus decay rate $\psi \rightarrow XX$ is given
by: \be \label{gamma} \Gamma_\psi  = { D}_\psi
\frac{m_\psi^3}{m_p^2}, \ee which reflects the fact that the
modulus is gravitationally coupled ($\Gamma_\psi \sim G_N \sim
m_p^{-2}$) and ${ D}_\psi$ is a model dependent constant that is
typically order unity. After the end of inflation, the Universe is
dominated by coherent oscillations of the inflaton field and $H
\sim \frac{2}{3t}$. After the decay of the inflaton and subsequent
reheating at temperature $T_r$,  the Universe is radiation
dominated and $H \sim \frac{1}{2t}$. In both these phases, the
evolution of the moduli can be written as: \ba\label{evolve}
\ddot{\psi} + {\cal O}(1)\frac{1}{t} \dot{\psi} + \frac{\partial
V}{\partial \psi}=0. \ea where we have neglected $\Gamma_{\psi}$
as it is planck suppressed. The minimum of the potential now is
{\it time-dependent} due to the time dependence of the Hubble
parameter. The evolution of the moduli in the presence of matter
and/or radiation as in the case above, has been studied in
\cite{Kaloper:1991mq,Brustein:2004jp,Brustein:1999yq,Huey:2000jx,Kaloper:2004yj,
hep-th/0403001,hep-th/0404177,Battefeld:2005av,hep-th/0601082,hep-th/0702220}.
In this case, as the modulus begins to roll down the potential, it
was shown in
\cite{Kaloper:1991mq,Huey:2000jx,Kaloper:2004yj,Battefeld:2005av})
that the presence of matter/radiation has a slowing effect on the
evolution of the field. This can naturally allow for the
relaxation of moduli into coherent oscillations about the
time-dependent minimum\footnote{We thank Joe Conlon and Nemanja
Kaloper for discussions on this approach.}. This `environmental
relaxation' can then slowly guide the modulus to the
time-dependent minimum.

Another possibility arises if the minimum of the potential is
located at a point of enhanced symmetry where additional light
degrees of freedom become important. This naturally arises in
SUGRA theories that are derived from string theories, where an
underlying knowledge of the UV physics is known
\cite{Dine:1998qr,hep-th/0403001,hep-th/0404177,Vafa:2005ui}. If the modulus initially has a
large kinetic energy, as it evolves close to the point of enhanced
symmetry, new light degrees of freedom will be produced and then
backreact to pull the modulus back to the special point of
enhanced symmetry. This simple example of `moduli trapping' is
present in a large number of examples in string theory with points
of enhanced symmetry \cite{hep-th/0403001,hep-th/0404177,hep-th/0601082,hep-th/0702220}.

The above effects lead to a natural solution of the so-called
`overshoot problem' \cite{Brustein:1992nk} (see also
\cite{Dine:1985he}), as argued below. As the universe expands and
cools, the Hubble parameter ($H$) decreases until it eventually
drops below the mass of the modulus $m_{\psi}$ ($\sim m_{3/2}$).
Thus, from (\ref{modulipotential}), we see that the first term in
the potential now becomes of the same order as the other two terms
and can no longer be neglected. At this time the modulus field
becomes under-damped and begins to oscillate freely about the true
minimum $\psi_0$ with amplitude $f_{\psi} \sim
(m_p^n\,m_{\psi})^{1/(n+1)}$. As an example, for $n=1$, $f_{\psi}$
is $~ (m_p\,m_{\psi})^{1/2}$ leading to a potential value $V \sim
m_{\psi}^2 f_{\psi}^2 \sim m_{\psi}^3m_p$ which is much smaller
than the overall height of the potential barrier at this time
($\sim m_{\psi}^2m_p^2$, as in any soft susy breaking potential).
Thus, there is no overshoot problem.

The modulus will now quickly settle into coherent oscillations at
a time roughly given by $t_{osc} =2H^{-1} \sim 2 m^{-1}_\psi$.
After coherence is achieved, the scalar condensate will then
evolve as pressure-less matter\footnote{If there are additional
terms that contribute to the potential (besides the soft mass),
then a coherently oscillating scalar does not necessarily scale as
pressure-less matter.}, i.e. $\rho_m(t_{osc}) = \half m_\psi^2
f_\psi^2$. Because the condensate scales as pressure-less matter
$\rho_m \sim 1/a^3$, its contribution relative to the background
radiation $\rho_r \sim 1/a^4$ will grow with the cosmological
expansion as $a(t) \sim 1/T$. Thus, if enough energy is stored
initially in the scalar condensate it will quickly grow to
dominate the total energy density.

\section{Overview of Results\label{overview}}

This section reviews the main results of the paper without 
technical details.

As explained above, the moduli start oscillating when the Hubble
parameter drops below their respective masses. Then they
eventually dominate the energy density of the Universe before
decaying. Within the context of $G_2$-MSSM models, the relevant
field content is that of the MSSM and $N+1$ real scalars. $N$ of
these are the moduli, $X_K$, of the $G_2$-manifold and the
remaining one is a scalar field, $\phi$, called the meson field,
which arises in the hidden sector dominating the supersymmetry
breaking. A reasonable choice for $N$ would be ${\cal O}(50)$ -
${\cal O}(100)$.

The masses are roughly as follows. The lightest particles beyond
the Standard Model particles are the gauginos. In terms of the
gravitino mass, $m_{3/2}$, their masses are of order
${\kappa}\,m_{3/2}$, suppressed by a small number $\kappa$.
$\kappa$ is determined by a combination of tree level and one-loop
contributions which turn out to be comparable.
The tree-level
contribution is suppressed essentially because $\phi$ dominates
the supersymmetry breaking, and to leading order, the gauge
couplings are independent of $\phi$. The precise spectrum of
gaugino masses is qualitatively similar, but numerically
different, to AMSB models. The LSP is a Wino in the $G_2$-MSSM,
similar to AMSB models. The current experimental limits on
gauginos require that the gravitino mass is at least 10 TeV or so.
In the $G_2$ framework, gravitinos naturally come out to be of
$\mathcal{O}(10-100)$ TeV \cite{hep-th/0701034}. 50 TeV is a
typical mass that we consider in this paper. The MSSM sfermions
and higgsinos have masses of order $m_{3/2}$, except the right
handed stop which is a factor of few lighter due to RG running. Of
the $N$ moduli, one, $X_N$ is much heavier than the rest, $X_i$.
The heavy modulus mass is about 600 $m_{3/2}$, while the $(N-1)$
light moduli are essentially degenerate with masses $\sim$
2$m_{3/2}$. Finally the meson mass is also about 2$m_{3/2}$. The
decays of the moduli and meson into gravitinos will therefore be
dominated by the heavy modulus $X_N$.

The decays can be parameterized by the decay width as, \be
\Gamma_X = D_X \frac{m_X^3}{ m_p^2} \ee reflecting the fact that
the decays are gravitationally suppressed. $D_X$ is a constant
which we calculate to be order one for the moduli but order 700
for $\phi$. So, the light moduli have decay widths of order
$10^{-13}$eV, corresponding to a lifetime of order $10^{-3}$ s.
The heavier scalars have shorter lifetimes, $10^{-5}$ s for $\phi$
and $10^{-10}$ for $X_N$, see tables 1 and 2. So, as the Universe
cools further and $H$ reaches a value of order $\Gamma_{X_N}$, the
heavy modulus decays. When this happens, the Universe is reheated
to a temperature, roughly of order $T_r \sim (\Gamma_{X_N}^2
m_p^2)^{1/4} \sim 40$ GeV. The entropy is increased in this phase,
by a factor of about $10^{10}$. This greatly dilutes the thermal
abundance of gravitinos and MSSM particles produced during
reheating (by the inflaton). The abundance of the light moduli and
meson are also diluted. Then, when $H$ reaches order
$\Gamma_{\phi}$ the meson decays. This reheats the Universe to a
temperature $T_r \sim 100$ MeV and increases the entropy by a
factor of order 100. Finally, as the Universe cools again and
reaches a temperature of about $10^{-13}$ eV the light moduli
decay. They reheat the Universe to a temperature of about 30 MeV
and a dilution factor of about 100 again. After this, all the
moduli have decayed and the energy density is dominated by the
decay of the light moduli. Since the final reheat temperature is
well above that of nucleosynthesis, BBN can occur in the standard
way.

Furthermore, since the entropy increases by a total factor of about $10^{14}$, the gravtino density
produced by moduli and meson decays is sufficiently diluted to an extent that it avoids existing bounds from
BBN from gravitino decays.

Since the energy density is dominated by the decaying light
moduli, the relic density of Wino LSP's is dominated by this final
stage of decay. The initial density of LSP's at the time of
production is such that the expansion rate is not large enough to
prevent self-interactions of LSP's. This is because
\be
\left. n^{initial}_{LSP} > \frac{3H}{\langle \sigma v \rangle}\right\vert_{T_r}
\ee where the
right side is to be evaluated at the final reheating temperature and $\sigma
v$ is the typical Wino annihilation cross-section  $\sim 10^{-7}
\mbox{GeV}^{-2}$. Therefore, the Wino's will annihilate until they
reach the density given on the R.H.S., which is roughly $10^{12}$ eV$^3$ - an energy density of $10^{23}$ eV$^4$. Here we have
assumed, as is reasonable, that since there is a lot of radiation
produced at the time of decay, the LSPs quickly become
non-relativistic by scattering with this `background'. Since the
entropy at the time of the last reheating $s \sim 10 \, T^3 \sim 10^{23}$ eV$^3$, the ratio of the energy density to entropy, is
around 1 eV. This should be compared to the observed value of this
ratio today, which is 3.6 $h^2$ eV, where the Hubble parameter
today is about 0.71.

Therefore, we see that the Wino LSP relic density is very reasonable in these models. The rest of this
paper is devoted to a much more precise, detailed version of this calculation.

\subsection{Scalar Decay and Reheating Temperatures}

Here we collect some more precise formulae for the decay and
reheat temperatures as a function of the moduli/meson masses.

The temperature at the time of decay can be found using \bea 3
H_d^2 &=& \frac{ m_\psi Y_\psi}{m_p^{2}} s_d =
\frac{m_\psi Y_\psi}{m_p^{2}} \left(\frac{2 \pi^2}{45}\right) g_{\ast s}(T_d) \; T_d^3, \\
&& \longrightarrow \; T_d = \left( \frac{30}{\pi^2}  \right)^{1/3}
\left( \frac{\Gamma_\psi^2 m_p^2}{m_\psi Y_\psi g_{\ast s}(T_d)}
\right)^{1/3}, \eea where $Y_\psi= n_\psi / s$ is the comoving
number density and \be \label{entropydensity}
s=\frac{\rho+p}{T}=\frac{2 \pi^2}{45} g_{*s} T^3, \ee is the
entropy density with $g_{*s}$ the number of relativistic degrees
for freedom\footnote{We will take $g_{*s}=g_*$, which is true if
all particles track the photon temperature. This is a good
approximation for most of the history of the universe (prior to
decoupling) \cite{Kolb:1990vq}.}. Parameterizing the decay rate as
above, i.e. $\Gamma_\psi= \cd_\psi m_\psi^3/m_p^2$ we find \be
\label{eq3} T_d = \left( \frac{30}{\pi^2} \right)^{1/3}
g^{-1/3}_{\ast s}(T_d) \left( \frac{\cd_\psi^2 m_\psi^5 }{ Y_\psi
m_p^2 } \right)^{1/3} \ee For later use we also note that if more
than one modulus dominates at the time of decay then the
temperature at the time of decay becomes \be T_d = \left(
\frac{30}{\pi^2} \right)^{1/3}  g^{-1/3}_{\ast s}(T_d) \left(
\frac{\cd_\psi^2 m_\psi^6 }{   m_p^2 \sum_i m_i Y_i }
\right)^{1/3} \ee where the sum is over all moduli (including the
one that decays). When the modulus decays, the relativistic decay
products will reheat the universe to a temperature, \bea
3H^2 &=& \frac{4 \Gamma_\psi^2}{3} =  m_p^{-2} \radw(T_r) \; T_r^4, \\
&& \longrightarrow \; T_r=\left( \frac{40}{ \pi^2} \right)^{1/4}
g_\ast^{-1/4}(T_r) \sqrt{\Gamma_\psi m_p}, \eea or \be
\label{eq1}
 T_r = \left( \frac{40}{\pi^2} \right)^{1/4} g_\ast^{-1/4}(T_r) \left(  \frac{\cd_\psi m_\psi^3 }{m_p} \right)^{1/2}.
\ee Instead, if more than one modulus contributes to the energy
density before decay the reheat temperature becomes \be
\label{treqn}
 T_r = \left( \frac{40}{\pi^2} \right)^{1/4} g_\ast^{-1/4}(T_r) \left(  \frac{m_\psi Y_\psi}{\sum_i m_i Y_i} \right)^{1/4}  \left(  \frac{\cd_\psi m_\psi^3 }{m_p} \right)^{1/2},
\ee where the sum is over all moduli (including the one that
decays) and we note that this could lead to a subdominant
radiation density compared to that of the remaining moduli. The
entropy production is characterized by (assuming that $\Delta \gg
1$) \be \label{eq2} \Delta=\left(  \frac{S_r}{S_d} \right) =
\frac{g_{\ast s}(T_r) a^3(t_r) T_r^3}{g_{\ast s}(T_d) a^3(t_d)
T^3_d}, \ee where $T_d$ and $T_r$ are the decay and reheat
temperatures, respectively. Making use of (\ref{eq1}),
(\ref{eq2}), and (\ref{eq3}) we find \bea
\Delta &=& \frac{2}{15}\left( 250 \pi^2  \right)^{1/4} \left(  \frac{g_{\ast s}(T_r) }{g_{\ast s}(T_d)} \right)  \left(  \frac{g_{\ast s}(T_d)}{g_\ast^{3/4}(T_r)} \right)  \frac{{m_\psi Y_\psi }}{\left( {\Gamma_\phi m_p}   \right)^{1/2}}, \nonumber \\
&=&   \frac{2}{15}\left( 250 \pi^2  \right)^{1/4}
g_\ast^{1/4}(T_r)  \left( \frac{m_p }{{\cd_\psi m_\psi}}
\right)^{1/2}Y_\psi, \label{usual2}. \eea For the case that more
than one modulus dominates the energy density before $\psi$
decays, we have instead \be \Delta=\frac{2}{15}\left( 250 \pi^2
\right)^{1/4}   g_\ast^{1/4}(T_r)  \left( \frac{m_p }{{\cd_\psi
m_\psi}}   \right)^{1/2} \left[ \frac{ \sum_i m_i Y_i}{m_\psi
Y_\psi} \right]^{1/4} Y_\psi, \label{bigdel} \ee where the sum
runs over all moduli that contribute to the energy density
(including the decaying modulus $\psi$).

\subsubsection{Moduli decay and BBN}
From (\ref{bigdel}), we see that the decay of moduli can produce a
substantial amount of entropy. Therefore, if any moduli present do
not decay before the onset of BBN the resulting entropy production
when decay occurs could result in devastating phenomenological
consequences.  However, another possibility is provided if the
late-time decay of the moduli reheat the universe to temperatures
greater than a few MeV.  Such reheating will then allow BBN to
proceed as usual.  Requiring that the modulus decay exceeds this
temperature one finds from (\ref{eq1}) that $m_\psi \gtrsim 10$
TeV.

\section{Summary of Results for the $G_2$-MSSM \label{reviewg2}}

In this section,
we give a brief summary of the results obtained in
\cite{hep-th/0606262,hep-th/0701034,Acharya:2008zi} for the $G_2$-MSSM.
Readers interested in more details should consult the references above.
$M$ theory compactifications on singular $G_2$ manifolds are
interesting in the sense that they give rise to $\mathcal{N}=1$
supersymmetry in four dimensions with non-Abelian gauge groups and
chiral fermions. The non-Abelian gauge fields are localized along
three-dimensional submanifolds of the seven extra dimensions
whereas chiral fermions are supported at points at which there is
a conical singularity. In order to study phenomenology concretely
one has to address the issues of moduli stabilization,
supersymmetry breaking and generation of the Hierarchy between the
Electroweak and Planck scales. These issues can be fairly successfully
addressed within the above framework.

In \cite{hep-th/0606262,hep-th/0701034,Acharya:2008zi}, it was
shown that all moduli can be stabilized generically in a large
class of $M$ theory compactifications by non-perturbative effects.
This happens in the zero-flux sector, our primary interest, when
these compactifications support (at least two) non-abelian
asymptotically free gauge groups. Strong gauge dynamics in these
non-abelian (hidden sector) gauge groups gives rise to the
non-perturbative effects which generate a moduli potential.
When at least one of the hidden sectors
also contains charged matter, under certain assumptions
defining the above framework,
supersymmetry is spontaneously broken in a metastable de Sitter
vacuum which is tuned to the observed value.
In the minimal framework, the
hidden sector, including its moduli and hidden sector matter, is
described by ${\cal N}=1$ supergravity with the following
K\"{a}hler potential $K$, superpotential $W$ and gauge kinetic
function $f$ at the compactification scale ($\sim M_{\rm unif}$):
\begin{eqnarray}
  K/m_p^2&=&-3\ln (4\pi ^{1/3}V_{7})+\bar{\phi}\phi, \quad V_7 =\prod_{i=1}^{N}
  s_i^{a_i}\nonumber\\
  W &=&m_p^3\left(C_1\,P\,\phi^{-(2/P)}\,e^{ib_1f_1}+C_2\,
Q\,e^{ib_2f_2}\right);
\;b_1=\frac{2\pi}{P},\,b_2=\frac{2\pi}{Q}\nonumber\\
f_1&=&f_2\equiv f_{\rm hid}=\sum_{i=1}^{N}\,N_i\,z_i; \;
z_i=t_i+is_i.
\end{eqnarray}
Here $V_{7}\equiv \frac{\mathrm{Vol}(X)}{l_{11}^7}$ is the volume
of the $G_{2}$ manifold $X$ in units of the eleven-dimensional
Planck length $l_{11}$, and is a homogenous function of the
$s_{i}$ of
degree $7/3$.
A simple and reasonable ansatz therefore is $V_{7}=\prod_{i=1}^{N}%
\,s_{i}^{a_{i}}$ with $a_{i}$ positive rational numbers subject to
the constraint $\sum_{i=1}^{N}a_{i}=\frac{7}{3}$.
$\phi\equiv\det(\mathcal{Q}\tilde{\mathcal{Q}})^{1/2}=(2\mathcal{Q}
\tilde {\mathcal{Q}})^{1/2}$ is the effective meson field (for one
pair of massless quarks) and $P$ and $Q$ are proportional to one
loop beta function coefficients of the two gauge groups which are
completely determined by the gauge group and matter
representations. The normalization constants $C_1$ and $C_2$ are
calculable, given a particular $G_2$-manifold. $f_{1,2}$ are the
(tree-level) gauge kinetic functions of the two hidden sectors
which have been taken to be equal for simplicity, (which is the
case when the corresponding two 3-cycles are in the same homology
class). $s_i$ are the $N$ geometric moduli of the $G_2$ manifold
while $t_i$ are axionic\footnote{These essentially decouple from
the moduli stabilization analysis. Hence they will not be
considered further.} scalars. The integers $N_i$ are determined from the
topology of the three-dimensional submanifold which supports the
hidden sector gauge groups.

If
volume of the submanifold supporting the hidden sector gauge
theories ($V_{\hat{Q}}$) is large, the potential can be minimized
analytically order-by-order in a $1/V_{\hat{Q}}$ expansion.
Physically, this expansion can be understood as an expansion in
terms of the small gauge coupling of the hidden sector -- $(\alpha
_{0})_{\mathrm{hid}}$, which is self-consistent since the hidden
sectors are assumed to be asymptotically free. The solution
corresponding to a metastable minimum with spontaneously broken
supersymmetry\ is given by
\begin{eqnarray}
s_{i} &=&\frac{a_{i}}{N_{i}}\frac{3}{14\pi }\frac{P_{\mathrm{eff}}\,Q}{Q-P}+%
\mathcal{O}(\P^{-1}),  \label{sol-1} \\
|\phi |^{2}
&=&1-\frac{2}{Q-P}+\sqrt{1-\frac{2}{Q-P}}+\mathcal{O}(\P^{-1}),
\label{sol-2}
\end{eqnarray}
where $P_{\mathrm{eff}}\equiv P\ln (C_{1}/C_{2})$. The natural
values of $P$ and $Q$ are expected to lie between ${\cal O}(1)$
and ${\cal O}(10)$. It is easy to see that a large
$P_{\mathrm{eff}}$ corresponds to small $\alpha $ for the hidden
sector
\begin{equation}
(\alpha _{0}^{-1})_{\mathrm{hid}}=\mathrm{Im}(f_{\mathrm{hid}})\approx \frac{%
Q}{2\pi (Q-P)}P_{\mathrm{eff}}
\end{equation}
implying that the expansion is effectively in
$P_{\mathrm{eff}}^{-1}$. The $\phi $ dependence of the potential
at the minimum is essentially
\begin{equation}
V_{0}\sim m_{3/2}^{2}M_{P}^{2}\left[ |\phi |^{4}+\left( \frac{4}{Q-P}+\frac{%
14}{P_{\mathrm{eff}}}-3\right) |\phi |^{2}+\left( \frac{2}{Q-P}+\frac{7}{P_{%
\mathrm{eff}}}\right) \right]
\end{equation}
Therefore, the vacuum energy vanishes if the discriminant of the
above expression vanishes, i.e. if
\begin{equation}\label{tune}
P_{\mathrm{eff}}=\frac{28(Q-P)}{3(Q-P)-8}
\end{equation}
The above condition is satisfied when the contribution from the
$F$-term of the meson field ($F_{\phi }$) to the scalar potential
cancels that from the $-3m_{3/2}^{2}$ term. In this vacuum, the
$F$-term of the moduli $F_{i}$ are much smaller than $F_{\phi }$.
Since phenomenologically interesting compactifications only arise
for $Q-P=3$ which corresponds to $\P=84$ from (\ref{tune}), we
will restrict our analysis to this particular choice.

\subsection{Moduli Masses}

Since in this paper we are interested in the evolution of the
moduli (and meson) fields, it is important to study their masses
in the vacuum described above. The set of gauge-singlet scalar
fields includes $N$ geometric moduli $s_i$ associated with $G_2$
manifold and a hidden sector meson field $\phi$. Since these
moduli and meson will mix in general, the physical moduli
correspond to mass eigenstates. The mass matrix can be written as:
\begin{eqnarray}
  \left(m_{X}^2\right)_{i\,j}&=&\left((a_i a_j)^{1/2}K_1 +\delta_{ij}
  K_2\right) \;m_{3/2}^2\\
  \left(m^2\right)_{i\,\phi}&=&(a_i)^{1/2}K_3\; m_{3/2}^2\\
  \left(m^2\right)_{\phi\phi}&=&K_4\; m_{3/2}^2
\end{eqnarray}
where $K_1$ to $K_4$ are obtained in \cite{hep-th/0701034}:
\begin{eqnarray}
  K_1&=&\frac{16}{9261}\left(\frac{Q}{Q-3}\right)^2\P^4\\
  K_2&=&
  \frac{22}{3}-\frac{8}{9\phi_0^2}-2\phi_0^2-(1+\frac{2}{3\phi_0^2})\frac{36}{\P}\\
  K_3&=&\sqrt{\frac{2}{3}}\left(\frac{16}{1323}\right)\left(\frac{Q}{Q-3}\right)^2\frac{\P^3}{\phi_0}\\
  K_4&=&\frac{32}{567}\left(\frac{Q}{Q-3}\right)^4\frac{\P^4}{\phi_0^2}
\end{eqnarray}
where $\phi_0^2\approx 0.734$. The special structure of the mass
matrix allows us to find the eigenstates analytically. There is
one heavy eigenstate with mass
$m_{X_N}=(7K_1/3+K_2)^{1/2}m_{3/2}$, $(N-1)$ degenerate light
eigenstates with mass $m_{X_j}=(K_2)^{1/2}m_{3/2}$ and an
eigenstate with mass
$m_{\phi}=(K_4-\frac{K_3^2}{K_1})^{1/2}m_{3/2}$. These mass
eigenstates of the moduli fields are given by:
\begin{eqnarray}
   X_{j}\,&=&\sqrt{\frac{a_{j+1}}{(\sum_{k=1}^{j} a_k)(\sum_{k=1}^{j+1}a_k)}}\;\left(\sum_{k=1}^{j}\sqrt{a_k}\,\delta
  s_k'
  -\frac{\sum_{k=1}^{j} a_k}{\sqrt{a_{j+1}}}\,\delta
  s_{j+1}'\right);\;j=1,2\cdots N-1\nonumber\\
  X_{N}&=& \sqrt{\frac{3}{7}}\sum_{k=1}^{N} \sqrt{a_k}\,\delta
  s_k'
\end{eqnarray}
where $\delta s_j'=\sqrt{\frac{3 a_j}{2s_j^2}}\;\delta s_j$ are
the canonically normalized moduli fields. The normalized moduli
fields can be related to the eigenstates by $\delta s_i'=U_{ij}
X_j$, in which $U_{ij}$ can be constructed using the eigenstates
listed above. It is easy to show that $(\vec X_j)_i=U_{ij}$ for
the eigenvector $\vec X_j$. In addition, there is another
eigenstate $X_{\phi}$ corresponding to the meson field. Actually,
the heavy eigenstate $X_N$ and $X_{\phi}$ mix with each other.
This mixing hardly changes the components of the eigenstate $X_N$
and $X_{\phi}$ since $m_{X_N} \gg m_{\phi}$. However, the mass of
the eigenstate $X_{\phi}$ is affected by the mixing. The masses
$m_{X_N}$ and $m_{\phi}$ only have a mild dependence on $Q$ (for
$\P=84,Q-P=3$), and do not depend on the number of moduli $N$ at
all. The mass of the light moduli $m_{X_j}$ does not even depend
on $Q$. Taking the expression for $K_2$, one immediately finds
that $m_{X_j} \approx 1.96\,m_{3/2}, \;j=1,\cdots,N-1$. This
result is very important since light moduli are then not allowed
to decay into gravitinos, essentially eliminating the moduli
induced gravitino problem. Choosing a reasonable value of $Q$ to
be of ${\cal O}(10)$, one finds that $m_{\phi}$ is roughly around
$2\,m_{3/2}$ while $m_{X_N}$ is roughly around $600\,m_{3/2}$.
Changing values of $Q$ by $\mathcal{O}(1)$ hardly changes the
moduli masses $m_{X_N}$ and $m_{\phi}$. Therefore, the above
typical values will be used henceforth in our analysis. To
summarize, the meson and moduli masses in the $G_2$ -MSSM
can be robustly determined in
terms of $m_{3/2}$.

\subsection{Couplings and Decay Widths}

Understanding the evolution of the moduli also requires a
knowledge of the couplings of the moduli (meson) fields to the
visible sector gauge and matter fields.
Since all the
moduli are stabilized explicitly in terms of the microscopic
constants of the framework, all couplings of the moduli and meson
fields to the MSSM matter and gauge fields can in principle be explicitly
computed. Here we focus on
the moduli couplings to MSSM matter and gauge fields. A different
visible sector, as might arise from an explicit construction, will
give rise to different couplings of the moduli fields in general,
although with roughly the same moduli masses.

Here we will give a brief account of the important couplings of
the moduli meson to visible gauge and matter fields and set the
notation. Details are provided in Appendix \ref{appendixa}. The
most important couplings of the moduli and meson fields involve
two-body decays of the moduli and meson to gauge bosons, gauginos,
squarks and slepton, quarks and leptons, higgses and higgsinos.
The three-body decays are significantly more suppressed and will
not be considered.

Let us start with the decay to gauge bosons and gauginos. The
relevant part of the Lagrangian is given by: \ba {\cal
L}_{\textrm{gauge boson, gaugino}} &=& g_{X_k
gg}\,X_k\,\hat{F}^a_{\mu\nu}\hat{F}^{a,\mu\nu}+
g_{X_k\tilde g\tilde g}\,X_k\,\hat{\lambda}^a\hat{\lambda}^{a}+\nonumber\\
& & g_{\hat{\delta\phi_0}\tilde g\tilde g}\,\hat{\delta
\phi_0}\,\hat{\lambda}^a\hat{\lambda}^{a};\;k=1\cdots
N;\;a=1,2,3\ea Here, $X_k$, $\hat{\delta \phi_0}$,
$\hat{F}^{a}_{\mu\nu}$ and $\hat{\lambda}^a$ are the normalized
moduli, meson, gauge field strength and gaugino fields
respectively. The expression for the couplings will be provided in
Appendix \ref{appendixa}. It is important to note that the meson
field does not couple to gauge bosons since the gauge kinetic
function $f_{sm}$ does not depend on $\phi_0$. The normalized
moduli eigenstates $X_k$ have already been discussed. The others
can be written as: \ba \hat{\delta \phi}_0 = \frac{\delta
\phi_0}{\sqrt{2}};\;
\hat{F}^{a}_{\mu\nu}=\frac{F^{a}_{\mu\nu}}{\sqrt{\langle{\rm
Im}(f_{sm})}\rangle};\;
\hat{\lambda}^a=\frac{\lambda^a}{\sqrt{\langle{\rm
Im}(f_{sm})}\rangle}\ea where $f_{sm}$ is the gauge kinetic
function for the visible SM gauge group. In the rest of the paper,
we will neglect the hats for these normalized fields and $m_p$ in
the couplings for convenience.

The coupling of the moduli and meson fields to the MSSM non-higgs
scalars (ie sfermions)  turn out to be important, as will be
seen later. Since the Standard model fermion masses (including that of the top)
are much smaller than that of the moduli, the decay of the moduli
and meson to these fermions will not be considered. The
coupling to the MSSM sfermions can be written as: \ba {\cal
L}_{\textrm{non-higgs scalars}} &=&
  (g'_{X\tilde f\tilde f})_{i,\alpha\beta}\left[\partial_{\mu}(X_i\;{\tilde
f}^{\alpha*})\partial^{\mu}{\tilde f}_{\alpha}\right] -
g^{\alpha}_{X\tilde f\tilde f}\,X_i\tilde{f}^{*\bar{\alpha}}\tilde{f}^{\alpha}\nonumber\\
&+&(g'_{\hat{\delta\phi}_0 \tilde f\tilde
f})_{i,\alpha\beta}\left[\partial_{\mu}(\delta\phi_0\;{\tilde
f}^{\alpha*})\partial^{\mu}{\tilde f}_{\alpha}\right] -
g^{\alpha}_{\delta\phi_0 \tilde f\tilde
f}\,\hat{\delta\phi}_0\tilde{f}^{*\bar{\alpha}}\tilde{f}^{\alpha}
\label{coup-geom-moduli3} \ea where $\tilde f_{\alpha}$ are the
canonically normalized scalar components of the visible chiral
fields $C_{\alpha}$, i.e. $\tilde f_{\alpha} =
\frac{C_{\alpha}}{\sqrt{\tilde{K}_{\alpha}}}$. The couplings to
the higgs and higgsinos are different due to the presence of the
higgs bilinear $Z\,H_uH_d + h.c$ in the K\"{a}hler
potential\cite{Acharya:2008zi}, which gives rise to contributions
to the $\mu$ and $B\mu$ parameters. In addition to the couplings
similar as those in Eq.(\ref{coup-geom-moduli3}), there are
additional couplings for scalar higgses, which can be
schematically written as: \ba {\cal L}_{\textrm{higgs}} &\supset&
g_{XH_dH_u}\,X_j H_d H_u+g'_{XH_dH_u}\,\partial_{\mu}X_j\partial^{\mu}(H_dH_u)+c.c\nonumber\\
&+&g_{\delta\phi_0H_dH_u}\,\delta\phi_0 H_d H_u+
g'_{\delta\phi_0H_dH_u}\,\partial_{\mu}\delta\phi_0\partial^{\mu}(H_d
H_u)+c.c \label{higgscoupling}\ea As explained in
\cite{Acharya:2008zi}, all higgs scalars except the SM-like higgs
and all higgsinos are heavier than the gravitino, implying that
the moduli and meson fields can only decay in this sector to the
light SM-like higgs ($h$). The coupling to the SM-like higgs can
be determined from the above coupling \label{higgscoupling} as
explained in appendix \ref{appendixa}.

Finally, the moduli and meson fields can also decay directly to
the gravitino. In fact, it turns out that the (non-thermal)
production of gravitinos from direct decays dominates the thermal
production of gravitinos in the early plasma. Therefore, it is
important to consider the moduli and meson couplings to the
gravitinos. Since the meson and light moduli are lighter than
twice the gravitino mass (as seen from the previous subsection),
only the heavy modulus can decay to the gravitino.

The explicit form of these couplings in terms of the microscopic
constants is provided in appendix \ref{appendixa}. An important
point to note is that these couplings are computed from the theory
at a high scale, presumably the unification scale. However, since
the temperature at which the moduli decay is much smaller than the
unification scale, one has to RG evolve these couplings to scales
at which these moduli decay (around their masses). The RG
evolution has also been discussed in appendix \ref{appendixa} for
the important couplings. Once the effective couplings of these
moduli and meson are determined, one can compute the decay widths,
as shown below.

For the $G_2$-MSSM model, we have found that light moduli and
meson dominantly decay to light higgses and squarks, while the
heavy modulus dominantly decay to light higgses only. In appendix
\ref{appendixa}, we have explicitly calculated the decay widths of
the moduli $X_k$ and meson. The widths of moduli can be
schematically written as:\ba \label{gamma1}\Gamma^{total}_{X_k}
\equiv \frac{D_{X_k}m_{X_k}^3}{m_p^2} &\approx&
\Gamma_{X_k\rightarrow gg} + \Gamma_{X_k\rightarrow \tilde g\tilde
g} + \Gamma_{X_k\rightarrow \tilde q \tilde
q}+\Gamma_{X_k\rightarrow hh}\\ &=& \frac{7}{72\pi}\left(N_G({\cal
A}_1^{X_k}+{\cal A}_{2}^{X_k})+{\cal A}_3^{X_k}+{\cal
A}_4^{X_k}\right)\,\left(\frac{m_{X_k}^3}{m_p^2}\right),\nonumber\ea
where $k=1\cdots N$ and $N_G=12$ is the number of gauge bosons or
gauginos. Note that ${\cal A}_{3}^{X_k}$ is significant only for
$k=1,2...,N-1$ (see appendix \ref{appendixa}). For the meson, the
width can be written as: \ba \label{gamma2}
\Gamma^{total}_{\hat{\delta \phi}_0} \equiv
\frac{D_{\phi}m_{\phi}^3}{m_p^2} &\approx&
\Gamma_{\delta\phi_0\rightarrow \tilde g\tilde g} +
\Gamma_{\delta\phi_0\rightarrow \tilde q\tilde q} + \Gamma_{\delta\phi_0\rightarrow hh}\nonumber\\
&=& \frac{7}{72\pi}(N_G{\cal A}_1^{\phi_0}+{\cal
A}_2^{\phi_0}+{\cal
A}_3^{\phi_0})\left(\frac{m_{\phi}^3}{m_p^2}\right).\ea

\subsection{Nature of the LSP}
Before moving on to discuss the evolution of moduli in the
$G_2$-MSSM, it is important to comment on the nature of LSP in
this framework. As explained in detail in \cite{Acharya:2008zi},
the $G_2$-MSSM framework gives rise to Wino LSPs for choices of
microscopic constants consistent with precision gauge unification.
Therefore, in our analysis we focus on the Wino LSP case. As we
will see, a Wino LSP turns out to be crucial in obtaining our
final result.

\section{Evolution of Moduli in the $G_2$-MSSM \label{sectionresults}} In this
section, we apply the general discussion in Section \ref{asection}
to the model of the $G_2$-MSSM reviewed in the previous section.
For clarity we will summarize our main results focusing on the
more salient aspects of the physics, leaving the more technical
details of the calculations to Appendix \ref{appendixb}.  We will
illustrate our computations with benchmark values, in order to get
concrete numerical results, and comment on the choice of the
benchmark values in section \ref{discussion}.

As discussed in Section \ref{asection}, we assume that
cosmological inflation and (p)reheating have provided adequate
initial conditions for the post-inflationary universe.

\subsection{Moduli Oscillations}
As reviewed in the last section, we have
a heavy modulus $X_N$, $N-1$ light moduli $X_i$, and
the scalar meson $\phi$. These will begin to oscillate in the
radiation dominated universe once the temperature cools and the
expansion rate becomes comparable to their masses.

For a benchmark gravitino mass value\footnote{We give detailed
numerical values for $m_{3/2}=50$ TeV.  It will be clear that
values a factor of two or so smaller or larger than this will not
change any conclusions in this and related analyses.} of $50$ TeV,
 the heavy modulus will begin
oscillations first, at around $t_{X_N}^{osc} \approx 10^{-32}$
seconds, corresponding to a temperature of roughly $T = 10^{12}$
GeV. Following the heavy modulus, the other moduli will begin
coherent oscillations around $10^{-30} \; s$ corresponding to a
temperature of roughly $10^{11}$ GeV. These results are summarized
in Table \ref{table1} below.
\begin{table}[htdp]
\begin{center}
\begin{tabular}{|c|l|l|}
\hline \;\; Modulus \;\; & \;\; Mass ($m_{3/2}=50$ \; TeV)  \;\; & \;\; Oscillation Time (seconds)  \;\;   \\
\hline   $X_N$  & \;\; $m_{X_N} = 600 \; m_{3/2}$ \;\; & \;\; $ t^{X_N}_{osc}=2 \times 10^{-32}$\;\;  \\
\hline   $X_\phi$  & \;\; $m_{\phi} \lesssim 2 \; m_{3/2}$ \;\; & \;\; $ t^{\phi}_{osc}=7 \times 10^{-30}$\;\;  \\
\hline   $X_i$ & \;\; $m_{X_i} \lesssim 2 \; m_{3/2}$ \;\; & \;\; $ t^{X_i}_{osc}=7 \times 10^{-30}$\;\;   \\
\hline
\end{tabular}
\end{center}
\caption{Oscillation times for the $G_2$-MSSM moduli}
\label{table1}
\end{table}

Since coherently oscillating moduli ($\rho_m$) scale relative to
radiation as $\rho_{m}/\rho_r \sim a(t) \sim 1/T$, the moduli will
quickly come to dominate the energy density of the universe, which is then matter dominated.
Following the beginning of coherent oscillations of the heavy
modulus, until the decay of all the moduli the universe will
remain matter dominated.  We will see that this, along with the
entropy produced during moduli decays, results in negligible
primordial thermal abundances of (s)particles compared with the
non-thermal abundances coming from direct decays of the moduli.
This will be crucial in addressing the gravitino problem and
establishing a Wino-like LSP as a viable dark matter candidate
through its non-thermal production.

\subsection{Moduli Decays and Gravitino Production}
As the universe continues to cool the expansion rate will
eventually decrease enough so that the moduli are able to decay.
This occurs when $H \sim \Gamma_X$, at which time the moduli will
decay reheating the universe and producing substantial entropy. We
will parameterize the decay rates of the $G_2$-MSSM moduli as: \be
\label{gamma2} \Gamma_X=\cd_X \frac{m_X^3}{m_p^2}, \ee
where $\Gamma_X$ is the decay width for particle $X$.  The decay
times will be computed for a set of benchmark values of $\cd_X$
for the various moduli (meson) which can be obtained by choosing
particular (reasonable) sets of values of the microscopic
constants (see appendix \ref{appendixa} for details).

\begin{table}[htdp]
\begin{center}
\begin{tabular}{|c|l|l|l|l|} \hline \;\; Modulus \;\;   \;\; & \;\; Decay constant  & \;\; Decay Time (seconds)  \;\; \\
\hline   $X_N$   & \;\; $\cd_{X_N}=2$\;\; & \;\; $ \tau_{X_N}= 9 \times 10^{-11}$ \;\; \\
\hline   $X_\phi$  & \;\; $\cd_{\phi}=710$\;\;  & \;\; $ \tau_{\phi}= 6 \times 10^{-6}$\\
\hline   $X_i$ & \;\; $\cd_{X_i}=4.00$ \;\;  & \;\; $ \tau_{X_i}= 10 \times 10^{-4}$\\
%\hline  \;\; $X_i$ (Model 2) \;\; & \;\; $  \cd_{X_i}^{II} ={7}/{(3\pi)}$ \;\;  & \;\; $ \tau_{X_i}= 5.23 \times 10^{-3}
%\left(  \frac{7/(3\pi)}{\cd_{X_i}^{II}} \right) \left(  \frac{50 \; TeV}{m_{3/2}} \right)^3 $\\
\hline
\end{tabular}
\end{center}
\caption{Decay constants and lifetimes for the $G_2$-MSSM moduli
for a set of benchmark microscopic values} \label{table2}
\end{table}

\subsubsection{Heavy Modulus Decay and Initial Thermal Abundances}
Given the $G_2$-MSSM values in Table \ref{table2} above,
the heavy modulus will be the first to decay at around
$10^{-11}s$. This decay will produce a large amount of entropy
$\Delta_{X_N}=S_{after}/S_{before} \approx 10^{10}$ (even though
the energy density of the heavy modulus is less than that of the
meson and moduli), reheating the universe to a temperature
$T_r^{X_N}=41$ GeV. The entropy production will not only dilute
the thermal abundances of all (s)particles, but also all the other
moduli. One particularly important non-relativistic decay product
of the heavy modulus is the gravitino.  Gravitinos will be
non-thermally produced by the modulus decay with a branching ratio
$B_{3/2}^{X_N}=0.07 \%$, which yields a comoving abundance
$Y_{3/2}^{(X_N)} = n_{3/2} / s \approx 10^{-9}$.  This can be
compared to the thermal abundance of gravitinos, which before
modulus decay is $Y_{3/2}^{thermal} = 2.67 \times 10^{-8}$. This
is further diluted by entropy production resulting from the decay,
i.e. $Y_{3/2}^{thermal} \rightarrow Y_{3/2}^{thermal} /
\Delta_{X_N} \approx 10^{-18}$. We see that the thermal
contribution to the gravitino abundance is negligible compared to
that from non-thermal production. A similar result
follows for all other (s)particles that are thermally populated
following inflation. Therefore, the primary source of
(s)particles, and in particular gravitinos and Lightest SUSY
Particles (LSPs), will result from non-thermal production
resulting from decays of the moduli.

\subsubsection{Meson/Light Moduli Decays and the Gravitino Problem}

The decay of the heavy modulus is followed by the decay of the
meson, at around $10^{-6}s$ (for benchmark values). The meson will
decay before the light moduli because of a larger decay width
compared to that for the light moduli (see appendix
\ref{appendixa} for details). Similar to the heavy modulus, the
meson contribution to the energy density is small compared to that
of the $N-1$ light moduli. Nevertheless, it produces some entropy
($\Delta_{\phi} \approx 121$) and reheats the universe to a
temperature of around $134$ MeV. The entropy production will again
dilute the abundance of light moduli, and any (s)particles
present, including the gravitinos from the heavy modulus decay.

The decay of the meson to gravitinos is particularly important, as
this can result in the well-known gravitino problem. If the scalar
decay yields a large number of gravitinos, these gravitinos can
later decay producing a substantial amount of entropy that could
spoil the successes of BBN.

The entropy produced from the decay of the meson and the
other light moduli further dilutes the gravitino abundance from
the heavy modulus.
The primary contribution to the gravitino
relic abundance comes from the decay of the heavy modulus since the
other fields have masses of order 2 $m_{3/2}$.
After
the decay of the meson, the energy density of the $N-1$ light
moduli is the dominant contribution to the total energy density of
the Universe.

Given that the $N-1$ light moduli are approximately degenerate in
mass, their decays will occur at nearly the same time, after the
decay of the meson. The resulting reheat temperature is found to
be approximately $32$ MeV, which is  an acceptable temperature for
consistency with the bound of $1$ MeV set by BBN
\cite{Kawasaki:1999na,Giudice:2000ex,Kawasaki:2000en,Hannestad:2004px}.

We note that the moduli decay rates have a strong dependence on
the gravitino mass (as it sets the moduli mass scale). So, the
decay of the light moduli being able to avoid BBN constraints is a
result of the fact that the gravitino mass is relatively large
($m_{3/2} \gtrsim 50$ TeV). However, as explained in detail in
\cite{Acharya:2008zi}, the gauginos are significantly suppressed
relative to the gravitinos allowing us to still obtain a light
($<$ TeV) spectrum which can be seen at the LHC. The decay of each
modulus will contribute to the total entropy production, and one
finds that the total entropy production for the set of benchmark
values of the microscopic constants is given by $\Delta_{X_i}=
418$. We also note that the light moduli lifetime depends
inversely on the decay constant ${D}_{X_i}$, so if instead of
taking relatively large values ${D}_{X_i}=4$ we take
relatively small values ${D}_{X_i}=0.4$, we find a reheat
temperature of $10$ MeV which is still compatible with
BBN\footnote{See appendix A for a discussion of the range of the
coefficients $\cd_{X_i}$.}. The decay of light moduli to
gravitinos is kinematically suppressed for the same reason as for
the meson. The final gravitino abundance is then just the
contribution from the heavy modulus decay diluted by the decay of
the meson and light moduli and is $Y_{3/2}^{final} =
Y_{3/2}^{\phi} / \Delta_{X_i} \approx 10^{-14}$ . The above
gravitino abundance is well within the upper bound on the
gravitino abundance set by BBN constraints, as it will not lead
to any significant entropy production at the time the gravitinos
decay. Thus, we find that there is {\em no gravitino problem} in
the $G_2$-MSSM. In addition to the relativistic decay products,
the light moduli will also decay appreciably into neutralinos
(LSPs), which we consider in detail in the next section.

\section{Dark Matter from the $G_2$-MSSM \label{darkmattersection}}
Natural models of electroweak symmetry breaking (EWSB) require
additional symmetries and particles beyond those of the Standard
Model.  The additional particles typically come charged under
additional discrete symmetries suppressing their decay to Standard
Model particles (e.g. R-parity, KK-parity, etc.), so such
models predict an additional, stable, weakly
interacting particle with an electroweak scale mass, i.e. they
naturally predict a candidate for Weakly Interacting Massive
Particle (WIMP) cold dark matter. In the case of the $G_2$-MSSM,
this gives rise to a Wino-like neutralino which is the lightest
supersymmetric particle (LSP) of the theory.

For completeness in section \ref{stdcdm} we will
review the standard calculation for computing the (thermal) dark
matter relic density today.  In section \ref{nstdcdm}, we will
then revisit this calculation for non-thermal production of LSPs
resulting from scalar decay.  In Section \ref{g2cdm}, we examine
how non-thermal production is naturally realized in the $G_2$-MSSM
and predicts the Wino LSP as a viable WIMP candidate.

\subsection{Standard Thermal Dark Matter \label{stdcdm}}
In the standard calculation of the relic abundance of LSPs it is
assumed that prior to BBN the universe is radiation dominated.  In
particular, it is assumed that the dark matter particles are
created from a thermal bath of radiation created from (p)reheating
after inflation. In this radiation dominated universe, the
Friedmann equation reads $3H^2=m_p^{-2} \rho_r$, with
$\rho_r=(\pi^2/30) g_* T^4$ the radiation density and $g_*$ the
number of relativistic degrees of freedom at temperature $T$.

The evolution of LSPs are given by the Boltmann equation
\be \label{bzeqn}
\dot{n}_X = -3 H n_X - \langle \sigma v \rangle \left[ n_X^2 - n_{eq}^2 \right],
\ee
where $\langle \sigma v \rangle$ is the thermally averaged cross-section, $n_X$ is the number
density, and $n_{eq}$ is the number density of the species in chemical equilibrium, i.e.
$XX \leftrightarrow \gamma \gamma$, where $\gamma$ is a relativistic particle such as the photon.

Assuming that initially the dark matter particles are relativistic
($m_X < T$) and in chemical equilibrium, then they will pass
through three phases as the universe expands and cools. Initially
their density will be determined by all the factors on the right
side of (\ref{bzeqn}). As long as the interactions of the
particles take place on smaller time scales than the cosmic
expansion then the particles will remain close to their
equilibrium distributions.  While the species is relativistic
($m_X<T$) this means that their comoving abundance is given by
$Y_X=n_X/s \approx Y_X^{eq} = const.$. Once the universe cools
enough from the cosmological expansion so that $X$ becomes
non-relativistic ($T<m_X$) then particle creation becomes more
difficult (Boltzmann suppressed) and the comoving abundance tracks
that of a non-relativistic species $Y_X \approx Y_X^{eq} = 0.145
\; x^{3/2} \exp(-x)$ where $x\equiv m_X/T$. The particle density
will continue to decrease until the number of particles becomes so
scarce that the expansion rate exceeds the annhiliation rate and
the particle species undergoes `freeze-out'.  From (\ref{bzeqn})
we see that at this time the number density is given by: \be
\label{fzeqn} n_X=\left. \frac{3 H}{ \langle \sigma v \rangle}
\right\vert_{T_f}, \ee where $T_f$ indicates that this relation
only holds at the time of freeze-out. Using (\ref{fzeqn}) and $Y_X
\approx Y_X^{eq}$ at the time of freeze-out, we find that
freeze-out is only logarithmically sensitive to the parameters of
the model, $x_f \equiv \frac{m_X}{T_f} \approx \ln \left[ m_X m_p
\langle \sigma v \rangle \right] $ and corrections are ${\cal O}(
\ln \ln x_f)$. Taking both the cross-section and mass $m_X$ to be
weak scale at around $100$ GeV we find that $x_f = 4$ and thus
the freeze-out temperature is $T_f= m_X/25 \approx 4$ GeV. From
(\ref{fzeqn}) and (\ref{entropydensity}), we find the comoving
density at freeze-out: \bea
Y_f&=& \frac{3 H}{s \langle \sigma v \rangle}, \\
&=& \frac{45}{2\pi \sqrt{10}} \frac{1}{\sigma_0g_*^{1/2}} \left( \frac{1}{m_p \langle \sigma v \rangle T_f} \right), \\
&=& \frac{45 }{2\pi \sqrt{10}} \frac{1}{\sigma_0g_*^{1/2}}  \left(
\frac{m_X}{ m_p} \right) x_f, \eea where we have taken $\langle
\sigma v \rangle = \sigma_0\,m_X^{-2}$.  We note that this answer
is rather insensitive to the details of freeze-out, and the
abundance is determined solely in terms of the properties of the
produced dark matter (mass and cross-section). In particular,
there is no dependence on the underlying microscopic physics of
the theory.

\subsection{Non-thermal Production from Scalar Decay \label{nstdcdm}}
We know from the successes of BBN that at the time the primordial
light elements were formed the universe was radiation dominated at
a temperature greater than around an MeV. However, perhaps
surprisingly, there is no evidence for a radiation dominated
universe prior to BBN. In particular, we have seen that in the
presence of additional symmetries and flat directions, scalar
moduli can easily dominate the energy density of the universe and
then later decay.  The presence of these decaying scalars can
alter the standard cold dark matter picture of the last section in
significant ways.

To understand this, consider the decay of an oscillating scalar
condensate $\phi$, which decays at a rate $\Gamma_\phi \sim
m_\phi^3/m_p^2$. When the expansion rate becomes of order the
scalar decay rate ($H \sim \Gamma$) the scalars will decay into
LSPs along with relativistic (s)particles which reheat the
universe to a temperature $T_r$. If this reheat temperature is
below that of the thermal freeze-out temperature of the particles
$T_f\sim m_X/25$ then the LSPs will never reach chemical
equilibrium. As an example, if we consider a scalar mass $m_\phi
\sim 10-100$ TeV this gives rise to a reheat temperature $T_r \sim
\sqrt{\Gamma_\phi m_p} \gtrsim$ MeV where $\Gamma_\phi \sim
m_\phi^3/ m_p^2$. The decay of $\phi$ in a supersymmetric setup
could lead to LSPs with weak-scale masses $m_X \sim$ 100 GeV,
which have a thermal freeze-out temperature $T_f\sim m_X / 25
\sim$ few GeV.  We see that in this case $T_r < T_f$ is quite
natural and the particles are non-thermally produced at a
temperature below standard thermal freeze-out. Thus, the particles
will be unable to reach chemical equilibrium.

Depending on the yield of dark matter particles from scalar decay, there are two possible outcomes of the non-thermally produced particles.

\subsubsection{Case one: LSP Yield Above the Fixed Point}
If the production of LSPs coming from scalar decay is large
enough, then some rapid annihilation is possible at the time of
their production. Since the particles are produced at the time of
reheating, we know from the Boltzmann equation (\ref{bzeqn}) that
the critical density for annihilations to take place is: \be
\label{nann} n_X^{c}=\left.\frac{3 H}{\langle \sigma v
\rangle}\right\vert_{T_r}, \ee which is \emph{different} from the
result (\ref{fzeqn}) in that here the reheat temperature and not
the thermal freeze-out is the important quantity. This is very
important because $T_r \sim \sqrt{\Gamma_\phi m_p}$ depends
on the microscopic parameters of the theory as the reheat
temperature is set by the decay rate of the scalar. In the
standard case, we saw that the freeze-out temperature, or more
precisely, the parameter $x_f\equiv m_X/T_f$ was only
logarithmically sensitive to the parameters of the dark matter and
gave no information at all about the underlying theory from which
the dark matter was produced (e.g. scalars from the underlying
microscopic physics).

Given that the initial number density of particles exceeds the
above bound ($n_X(0)> n_X^{c}$), the LSPs will quickly annihilate
until they reach the density (\ref{nann}).  Thus, the critical value $n_X^c$ serves as a fixed point
for the number density, since any production above this limit will always result in the same
yield of particles given by $n_X^c$.
From this one finds the
comoving density \cite{Moroi:1999zb}
\be Y_X=\frac{c_1}{g_*^{1/2}}\frac{1}{m_p
\langle \sigma v \rangle T_r}= Y_X^{std} \left(
\frac{T_f}{T_r}\right), \ee where $c_1=45/(2\pi \sqrt{10})$. We
see that non-thermal production can yield a greater comoving
density than standard thermal production by a factor ($T_f/T_r$).
For the example considered above, namely $m_\phi \sim 10-100$ TeV,
$m_X \sim$ 100 GeV, and $T_r \sim$ few MeV we find the comoving
density is enhanced by a factor $\sim 10^2-10^3$. One interesting
consequence of this is it allows room for larger annihilation
cross-sections for the LSPs.  For example, in standard thermal
production a Wino-like LSP leads to too small a relic density
since its annihilation cross section is only s-wave suppressed .
In the case of the $G_2$-MSSM, non-thermal production is a natural
consequence of the microscopic physics and a Wino LSP will provide
a perfectly suitable WIMP candidate.

\subsubsection{Case two: LSP Yield Below the Fixed Point}
The other possibility is that the decay of the scalar yields few enough LSPs ($n_X(0) < n_X^c$) so that annihilation does not occur.
Then the comoving abundance is simply given by
\be
Y_X= B_\phi \Delta_\phi^{-1} Y_\phi^{(0)} \sim \frac{B_\phi n_\phi^{(0)}}{T_r^3},
\ee
where $B_\phi$ is the branching ration of scalars to LSPs and $Y_\phi^{(0)}$ is the initial abundance of scalars in the decaying condensate.
We note that again this result depends on the underlying physics of the UV theory, since both the branching ratio and the reheat temperature are coming from the physics
of the scalar.

\subsection{Dark Matter in the $G_2$-MSSM \label{g2cdm}}
As shown in \cite{Acharya:2008zi}, the LSP in the $G_2$-MSSM is
predominantly Wino-like. There are two significant sources of
these LSPs in the $G_2$-MSSM -- direct production from decays of
both the gravitino and the light moduli. As explained earlier, the
thermal abundance of LSPs in the early plasma after inflation is
vastly diluted by the entropy productions from the heavy modulus,
meson and the light moduli. Therefore, the thermal abundance of
LSPs is negligible. In addition, the LSPs produced from decays of
the heavy modulus and the meson field are also diluted by the
entropy production from the light moduli and are negligible as
well.

The light moduli may decay to LSPs directly, or via decay to superpartners. From Section
\ref{reviewg2} the branching ratio for this process to occur for a
set of benchmark values of the microscopic paramaters is
$B_{LSP}^{X_i} \sim 25 \%$ and the comoving abundance is then
found to be: \be \label{LSPdensity}
Y_{LSP}^{(X^{i})}=\Delta_{X_i}^{-1} B_{LSP}^{X_i} (N-1)
Y_{X_i}^{(\phi)} =1.19 \times 10^{-7}, \ee where $\Delta_{X_i}=
417.7 \; \left[(N)/100 \right]^{3/4}$ is the entropy production
from the decay of all the light moduli $X_i$. Here we have taken
benchmark value for the number of light moduli to be $100$. The
corresponding number density at the time of reheating is \bea
n_{LSP}&=&s(T^{X_i}_r) Y_{LSP}, \\
&=& 1.79 \times 10^{-11} \; \mathrm{GeV^3} \label{lsps} \eea As
discussed in the last section, we must compare this number density
of LSPs to that of the critical density for annihilations
(\ref{nann}). At the time of reheating from the light moduli the
Hubble parameter is given by \be H(t_r)=\left( \frac{\pi^2
g_\ast}{90} \right)^{1/2} \frac{(T_r^{X_i})^2}{m_p}=4.48 \times
10^{-22} \; \mathrm{GeV}. \ee The dominant (s-wave) annihilation
cross section for the LSPs ($\tilde{W}^0 \tilde{W}^0 \rightarrow
W^+ W^-$) is given by \be \label{cs}\langle \sigma v \rangle=
\sigma_0\,m_{LSP}^{-2} = \frac{1}{m_{LSP}^2}\,\frac{g_2^4}{2 \pi}
\frac{(1-x_w)^{3/2}}{(2-x_w)^2} = 3.26 \times
10^{-7}\;\mathrm{GeV^{-2}}, \ee where $x_w=m_w^2/m_{LSP}^2$,
$m_w=80.4$ GeV is the $W$-boson mass, and $g_2\approx 0.65$ is the
gauge coupling constant of $SU(2)_L$ at temperatures $T_r \sim$
MeV, and this defines $\sigma_0$. It is crucial that the cross-section is s-wave so that there
is \emph{no} temperature dependence in $\langle \sigma v \rangle$.
We will comment more on this in section \ref{discussion}. Using
(\ref{nann}) we find the fixed point density for annihilations \be
\label{criticaldm} n_{LSP}^c=4.12 \times 10^{-15} \;
\mathrm{GeV^3}. \ee We see that the produced density is greater than
the fixed point value $n_{LSP}^{X_i} > n_{LSP}^c$ and annihilations
will occur. This corresponds to the ``LSP yield above the fixed point"
case discussed above. Thus, the LSPs produced will quickly
annihilate down toward the fixed point value in less than a Hubble time.
The relic density of dark matter is then given by the fixed point
value (\ref{criticaldm}) and the critical density of LSPs today
coming from decay of the light moduli is \bea
\Omega_{LSP}^{X_i}=\frac{m_{LSP} Y^c_{LSP}}{\rho_c/s_0}=
\frac{1}{\rho_c/s_0}\,\left(\frac{45}{2\pi\sqrt{10g_*}\sigma_0}\right)\left(\frac{m_{LSP}^3}{m_p\,T_r^{X_i}}\right)
=0.76 \; h^{-2} \eea where $s_0$ and $\rho_{c}$ are the entropy
density and critical density today, respectively, and we have used
the experimental value $\rho_c / s_0= 3.6 \times 10^{-9} h^2 \;
GeV$ with $h$ parameterizing the Hubble parameter today with
median value $h=0.71$.

In addition to this contribution, there is also the contribution
from the decay of non-thermal gravitinos produced from the heavy
modulus which have a final abundance $Y_{3/2}^{final} \approx
10^{-14}$. The contribution from gravitinos to the critical
density of dark matter is then \be
\Omega_{LSP}^{(3/2)}=\frac{m_{LSP} Y_{3/2}^{final}
s_0}{\rho_c}=0.0008 h^{-2}, \ee which is negligible compared with
that coming from the light moduli.

Thus, the total critical density in dark matter coming from the
LSPs of the $G_2$-MSSM is :{\footnotesize \be \Omega_{LSP}\,h^2
\approx 0.27 \left( \frac{m_{LSP}}{100 \, \mathrm{GeV}} \right)^3
\left( \frac{10.75}{g_\ast(T_r)} \right)^{1/4} \left( \frac{3.26
\times 10^{-7} \mathrm{GeV^{-2}}}{\langle \sigma v \rangle}
\right) \left( \frac{4}{\cd_{X_i}} \right)^{1/2} \left( \frac{2 \,
m_{3/2}}{m_{X_i}} \right)^{3/2} \left(  \frac{100 \;
\mathrm{TeV}}{m_{3/2}} \right)^{3/2}, \label{LSPlight} \ee } where
we have included all the parametric dependence of the answer
derived in Appendix B. This value should be compared to the
experimental value $\Omega_{CDM} h^2 = 0.111 \pm 0.006$
\cite{Komatsu:2008hk}. For those used to $\langle \sigma v
\rangle$ in other units, note that $1 \, \mathrm{GeV}^{-2} = 0.4
\times 10^{-27} cm^2$.

This result is not presented in terms of central values -- rather
it is the best value we can obtain. The LSP mass can be larger
than $100$ GeV, but not smaller.  The decay constant $D_{X_i}$ can
be order 4, but a scan of the microscopic parameter space suggests
a somewhat smaller value for the only calculable example so far
known (see appendix B.4).  A better understanding of the string
theory could give 4 or a larger value. Whereas $m_{3/2}$ is
somewhat constrained to be at most about $100$ TeV by the
parameters of the framework, as explained in
\cite{Acharya:2008zi}. Therefore, this framework is rather
constrained and predictive. We view the closeness of this result
as a success, and as an indication that improving the underlying
theory may improve the agreement with data.

\section{Discussion of Results}\label{discussion}
We have seen in the previous sections that for natural values of
microscopic parameters, there is no moduli and gravitino probem in
realistic $G_2$ compactifications. In addition, within the
$G_2$-MSSM, the non-thermal production of Wino LSPs from the light
moduli give rise to a relic density with the right order of
magnitude (up to factors of a few). It is possible that with a
more sophisticated understanding of the theory, one could obtain a
result more consistent with the observational results. It is also
worthwhile to understand these results from a physical point of
view. The results obtained above depend surprisingly little on
many of the details of the microscopic parameters. In particular,
there is essentially no dependence of the final relic density on
the total number of moduli ($N$), the masses ($m_{X_N},m_{\phi}$)
and couplings ($D_{X_N},D_{\phi}$) of the heavy modulus and meson
fields as well as the initial amplitudes of the moduli ($f_{X_k}$)
and meson ($f_{\phi}$) fields. This is good in a sense since our
understanding of the underlying theory and many of the above
microscopic parameters is incomplete. However, the result
\emph{does} depend crucially on certain qualitative (and also some
quantitative) features of the underlying physics, as we discuss
below.  In general it is better if results depend on the
microscopic theory, since then data can tell us about the
underlying theory.

One very important feature which helps avoid the gravitino problem
is that the meson and light moduli have masses which are of order
(actually slightly below) two gravitino masses,
as we saw explicitly in Section 4.1, 
This kinematically suppresses their decays to
the gravitino.
The gravitino abundace is thus dominated by
decay of the heavy modulus which is further diluted by entropy
production from the decays of the meson and light moduli.
Therefore, a natural mechanism for solving the gravitino problem
in a generic setup is that the modulus which decays last does not
decay to the gravitino, The moduli problem can also be easily solved in
frameworks where the gravitino mass is $\gtrsim 10$ TeV, which is
naturally satisfied in the $G_2$ framework.

Another qualitative feature of the $G_2$ framework is that there
is a hierarchy in the time scales of decay of the various moduli
(meson) fields. Since the mass of the heavy modulus is much larger
($\sim$ 300 times) than that of the other moduli (meson), it decays much
earlier. Also, from our current understanding of the K\"{a}hler
potential of the meson and moduli fields, one finds (see appendix
\ref{appendixa}) that the meson decays before the light moduli due
to a larger decay width. The precise computation of the decay
width depends on the nature of the K\"{a}hler potential for the
meson and moduli and the K\"{a}hler metric for matter fields, and
one might argue that there are inherent uncertainties in our
understanding of these quantities. However, 
the only qualitative feature relevant for cosmological
evolution is that the meson decays {before} the light moduli.
As long as the light moduli decay {last} (which we have
argued in the appendix to be the natural case from our current
understanding of the K\"{a}hler potential), the result does
{not} depend on any of the masses and couplings of the heavy
modulus or the meson field. The final result depends only on the
masses and couplings of the light moduli which decay {last}.
The same qualitative feature could be present in other frameworks
arising from other limits of string/$M$ theory.

Now that it is clear that it is the light moduli decaying at the
end which affect the final relic density, it is important to
understand their effect more closely. In any theory of (soft)
supersymmetry breaking, the
mass of the light moduli will be set by the gravitino mass scale.
In the context of low energy supersymmetry, the gravitino mass
will typically be in the range $1-100$ TeV. Therefore, the light
moduli will also be typically in the above range\footnote{This is
however not true for Large Volume compactifications as the
lightest modulus in that case is much lighter than $m_{3/2}$ \cite{hep-th/0505076}.}.
Since the reheat temperature of the moduli basically depends on
the moduli masses (assuming the coefficient $D_{X_i}$ is
$\mathcal{O}(1)$), the light moduli will typically give rise to a
reheat temperature $T_r^{X_i}$ of $\mathcal{O}(1-100)$ MeV, which
is far smaller than the freezeout temperature of the LSPs ($T_f^{LSP}
\sim$ GeV) which could be produced from the light moduli. This is
true for the $G_2$ framework and could be true for many other
frameworks as well. Therefore, with $T_r^{X_i} < T_f^{LSP}$, the
final outcome for the relic density will depend on the whether the
number density of the LSPs produced from the light moduli
($n_{LSP}^{(X_i)}$) is greater or smaller than the critical number
density at $T_r^{X_i}$ ($n_{LSP}^{(c)}|_{T_r^{X_i}}$).

For the $G_2$ framework, for natural values of the microscopic
parameters one finds that $n_{LSP}^{(X_i)} >
n_{LSP}^{(c)}|_{T_r^{X_i}}$ as shown in section \ref{g2cdm}. This
is equivalent to the inequality:\bea
\label{constraint}B^{(X_i)}_{LSP}\,D_{X_i} &>&
\frac{1.5}{\sigma_0}\,(\frac{m_{LSP}^2}{m_{X_i}^2}) \approx
120\,\gamma^2\nonumber\\ \mathrm{with}\; \gamma &\equiv&
\frac{m_{LSP}}{m_{3/2}}\eea where $\sigma_0$ is defined by
(\ref{cs}) and we have used $m_{X_i}\approx 1.96\,m_{3/2}$. As
explained in \cite{Acharya:2008zi}, the quantity $\gamma$ depends
predominantly on $\delta$, which characterizes the threshold
correction to the gauge couplings at the unification scale. The
dependence on other microscopic parameters such as $V_7$ and $C_2$
(see section \ref{reviewg2}) is largely absorbed into the
gravitino mass. The suppression factor $\gamma$ depends almost
linearly on $|\delta|$, and typically takes value in the range
$\sim (1-6)\times 10^{-3}$. Now, the constraint (\ref{constraint}) is
easy to understand. For natural values of microscopic parameters
in the $G_2$ framework, one has $B_{LSP}^{(X_i)}={\cal O}(25\%)$,
$D_{X_i}={\cal O}(1)$ (see appendix \ref{appendixa}) which easily
satisfy (\ref{constraint}) above. In order for other frameworks to
realize this situation, a criterion similar to (\ref{constraint})
needs to be satisfied.

When (\ref{constraint}) is satisfied, the final relic density can
be written as (see (\ref{criticaldm}) and (\ref{LSPlight})): \bea
\label{final} \Omega_{LSP}\,h^2&\approx&\frac{m_{LSP}
Y^c_{LSP}}{\rho_c/s_0}=
\frac{1}{\rho_c/s_0}\,\left(\frac{45}{2\pi\sqrt{10g_*}\sigma_0}\right)\left(\frac{m_{LSP}^3}{m_p\,T_r^{X_i}}\right)\nonumber\\
&\approx&\frac{1}{\rho_c/s_0}\,\left(\frac{45}{2\sqrt{10\pi}(40g_*)^{1/4}\sigma_0}\right)\left(\frac{m_{LSP}^3}
{D_{X_i}^{1/2}m_p^{1/2}m_{X_i}^{3/2}}\right)\nonumber\\
&\approx& 18
\;\mathrm{GeV}^{-3/2}\left(\frac{m_{LSP}^{3/2}\gamma^{3/2}}{D_{X_i}^{1/2}}\right)=
18\;\mathrm{GeV}^{-3/2}\left(\frac{m_{3/2}^{3/2}\gamma^{3}}{D_{X_i}^{1/2}}\right)\eea
An upper bound on the observed value of the relic density
implies that smaller values of $\gamma$ and $m_{LSP}$ and larger
values of $D_{X_i}$ are preferred. A small $\gamma$ implies that
for a given LSP mass a heavier gravitino is preferred implying
that the moduli be correspondingly heavier. Also, since $\gamma$
is roughly linear in $|\delta|$, smaller values of $|\delta|$ are
preferred. These features can be seen easily from the plots in
figures \ref{oh2-contour} and \ref{reheat}.
Figure \ref{oh2-contour} shows a contour plot of the relic density in the
$D_{X_i}-m_{3/2}$ plane for two (large and small) values of
$|\delta|$ which correspond to two (large and small) values of
$\gamma$.
\begin{figure}[h!]
    \begin{tabular}{c}
      \includegraphics[width=12cm,angle=0]{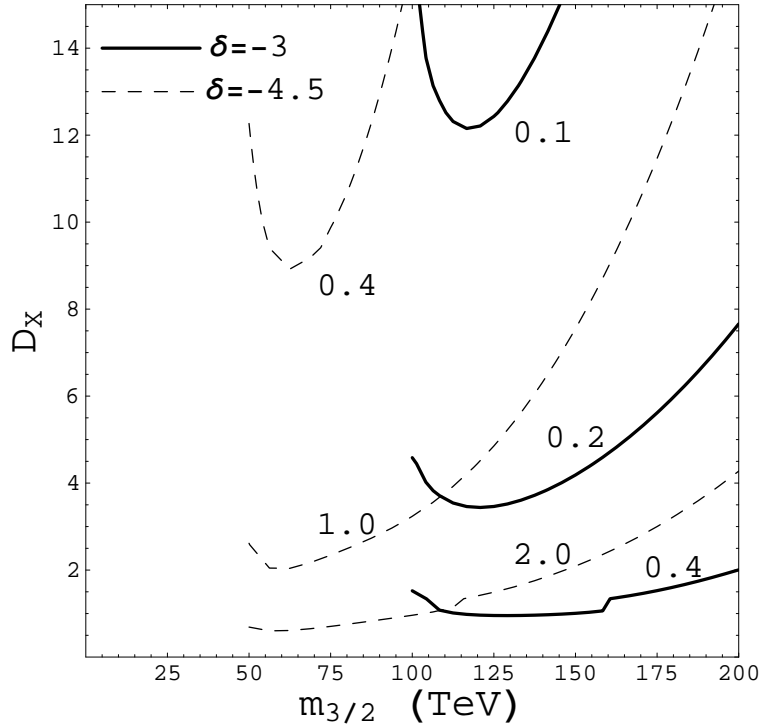}
    \end{tabular}
    \caption{The contour plot of the relic density in the $G_2$-MSSM in the $D_{X_i}-m_{3/2}$ plane for two (large and small) values of
$|\delta|$ which correspond to two (large and small) values of
$\gamma$. The solid lines are for $\delta=-3$ (a correction to
$\alpha^{-1}_{unif}$ of order $3/26$), and the dashed lines for
$\delta=-4.5$.}
    \label{oh2-contour}
\end{figure}
\begin{figure}[h!]
    \begin{tabular}{c}
     \includegraphics[width=12cm,angle=0]{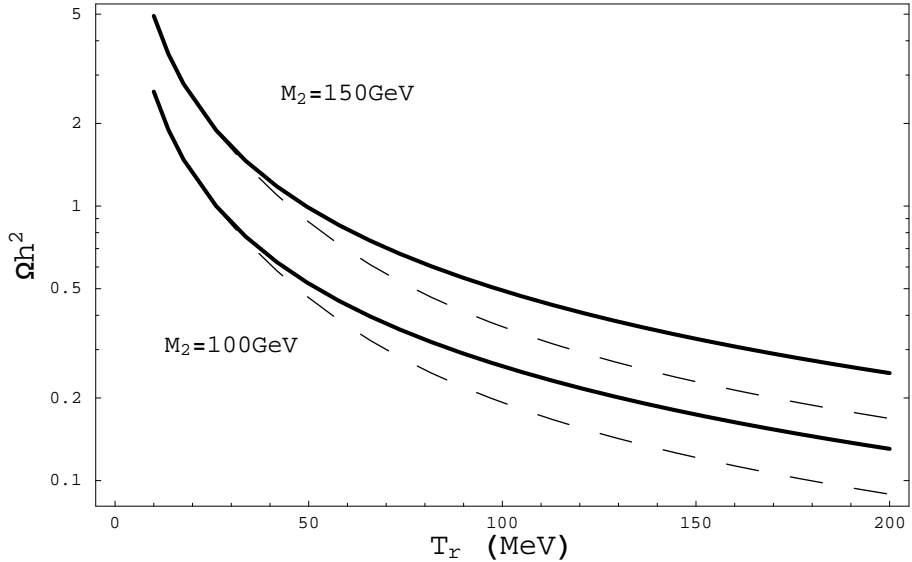}
    \end{tabular}
    \caption{The LSP relic density for the $G_2$-MSSM plotted as a
function of the reheat temperature of the light moduli.
    The solid line assumes no coannihilation with charged Winos; the dashed line includes
coannihilation with charged Winos.}
    \label{reheat}
\end{figure}
Figure \ref{reheat} shows the dependence of the relic density on
the reheat temperature of the light moduli ($T_r^{X_i}$). As seen
from the first line in (\ref{final}), the relic density is
inversely proportional $T_r^{X_i}$ implying that a higher reheat
temperature is preferred. A higher $T_r^{X_i}$ corresponds
precisely to a larger $D_{X_i}$ and $m_{X_i}$ (larger $m_{3/2}$)
as explained above.

As explained in section \ref{g2cdm}, the nature of the LSP is also
crucial to the final result for the relic density. For the $G_2$
framework, the annihilation cross-section is s-wave and does not
depend on $T_r^{X_i}$. On the other hand, if the LSP were Bino,
the cross-section would be p-wave suppressed and would depend
linearly on $T_r^{X_i}/m_{LSP}$, thereby making it suppressed
relative to the s-wave result. This would make the relic density
much larger than the result obtained for the s-wave case above.
Therefore, the upper bound on relic density prefers small mixing
angles (or vanishing mixing angles, as in the $G_2$-MSSM) with the
Bino and Higgsino components. This can be seen from figure
\ref{mixing}.
\begin{figure}[h!]
    \begin{tabular}{c}
     \includegraphics[width=12cm,angle=0]{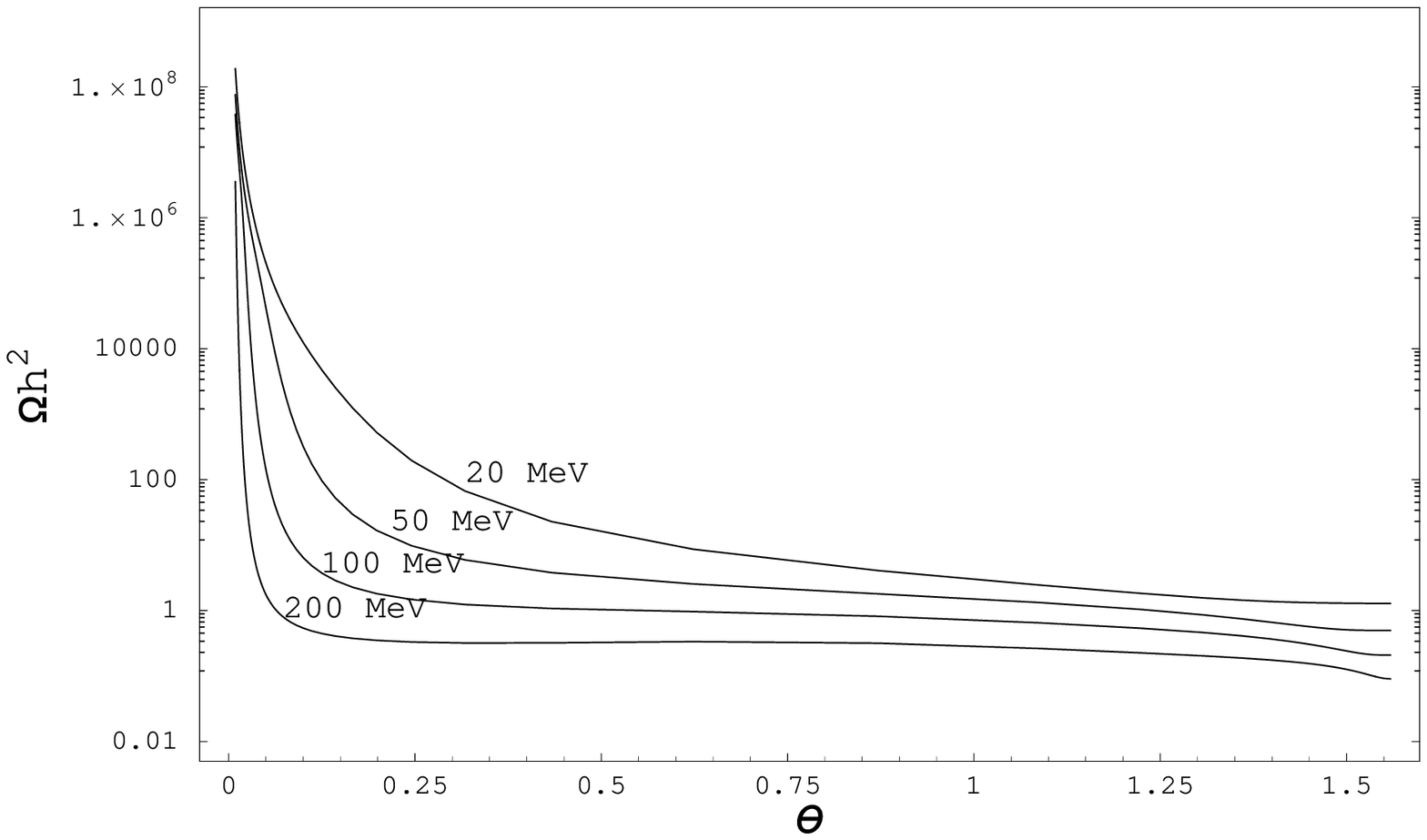}
    \end{tabular}
    \caption{The LSP relic density in the $G_2$-framework plotted
as a function of the mixing angle of Bino and Wino for
$M_2=100$GeV. }
    \label{mixing}
\end{figure}

\section{Summary and Future Directions}

In this paper we have emphasized the importance of the
cosmological moduli and gravitino problems and the relation to
adequate generation of dark matter in thermal equilibrium, or generation
of too much dark matter non-thermally in string/$M$ theory
frameworks. Focussing on $G_2$ compactifications, in
particular on the $G_2$-MSSM, we have found that the decay of
moduli in this framework is rather naturally consistent with BBN
constraints, and the associated large entropy production at late
times (but before BBN) results in an avoidance of the gravitino
problem(s). Moreover, we have seen that the late decay of the
light moduli into Wino-like neutralinos leads to a nearly acceptable
relic density of cold dark matter. This result arises from a
combination of entropy production and LSPs from moduli decay giving
an adequate relic density from non-thermal production of dark
matter.  This process offers an explicit example of how {\it thermal dark matter
production is not the dominant source of cosmological dark
matter, especially in the presence of moduli.}
The LSP is Wino-like here as well as in anomaly
mediated theories, but for interestingly different reasons --
here the tree level gaugino masses are universal but about the same size as the anomaly
mediated ones, and the finite one loop Higgsino is comparable with both.

The result for the final relic density depends parametrically on
the couplings and mass of the light moduli (which decay last) and
the mass of the LSP. The masses of the light moduli and the LSP
are set by the gravitino mass scale and depend on a set of
underlying microscopic parameters of the theory. The couplings of
the moduli depend on the K\"{a}hler potential of the theory. Since
our understanding of the K\"{a}hler potential is incomplete, it is
only possible to make reasonable assumptions to proceed, which is
what we have done, but one can see that most of the results are insensitive to these uncertainties.  
That is because the moduli decays produce a large number density of LSPs, which then
annihilate down to the final relic density that only depends on the reheating temperature.
From (\ref{final}) and figure
\ref{oh2-contour}, we see that an upper bound on the relic density
prefers a light LSP, heavy gravitino and large couplings to the
visible sector parameterized by $D_{X_i}$ (defined in
(\ref{gamma1})). These results obtained have been explained in
terms of the underlying qualitative features of the framework.
These qualitative features could be present in other string/$M$
theory frameworks as well, leading to similar results.

There is not yet a satisfactory inflation mechanism for the $G_2$-MSSM.
This is under study.  Fortunately, our results are not sensitive to that.
We assume only that at an early time inflation ends and the 
energy density of the universe is dominated by moduli settling into the minimum of the potential.

In future, one would like to understand the origin of the baryon
asymmetry in the Universe (BAU) within string/$M$ theory
frameworks. In the $G_2$ framework, the large entropy production
resulting from the decay of the moduli was crucial for addressing
the gravitino problem. However, this entropy will also act to
reduce any initial baryon asymmetry. Therefore, one requires a
large initial asymmetry or a late-time mechanism for regeneration
of the asymmetry. For example, a large initial baryon asymmetry
could arise from the Affleck-Dine mechanism \cite{Affleck:1984fy}, or it could happen
that the superpartner parameter space allows for late-time
electroweak baryogenesis. This is work in progress.

Understanding the above issues would be crucial to solving the
``cosmological inverse problem" (see
\cite{Simon:2006du,Balasubramanian:2007nu} for some preliminary
work in this direction), usually considered separate from the
``LHC Inverse Problem" \cite{ArkaniHamed:2005px}. Within the context of realistic
string/$M$ theory frameworks, however, the two inverse problems
merge into one ``inverse problem" as the microscopic parameters
characterizing the underlying physics of any framework have
predictions (at least in principle) for both particle physics as
well as cosmological observables, thereby providing unique connected
insights into these basic issues.

\section{Acknowledgements}
We thank Lotfi Boubekeur, Joe Conlon, Paolo Creminelli, Sera Cremonini, Alan Guth, Lawrence Hall, Nemanja Kaloper, 
Joern Kersten, Siew-Phang Ng, 
Piero Ullio, Filippo Vernizzi, and Lian-Tao Wang for useful discussions.  
The research of K.B., G.L.K., P.K., 
J.S. and S.W. is supported in part by the US Department of Energy.
S.W. would also like to thank the MIT Center for Theoretical Physics for hospitality.
J.S. would also like to thank the Institute for Advanced Study - Princeton for hospitality.
\appendix

\section{Cosmology of the $G_2$-MSSM Moduli -- A detailed treatment \label{appendixb}}
In this appendix, we include detailed calculations leading to the
abundances, entropy production, and reheat temperatures quoted in
the paper for sets of benchmark values of the microscopic
parameters. The computation of couplings and decay widths of the
moduli and meson fields in terms of the microscopic parameters
which motivate the benchmark values will be given in appendix
\ref{appendixa}. We have retained the parametric sensitivity to
the gravitino mass, number of moduli (topology), and the overall
couplings of the moduli (meson) in order to address the robustness
and plausibility of the framework.

\subsection{Heavy modulus oscillations}
At the time the heavy moduli ($X_N$) starts coherent oscillations
the universe is radiation dominated and the Hubble equation is
given by \be 3 H^2 =3 \left( \half m_{X_N}\right)^2= m_p^{-2}
\rad. \ee The temperature at which the modulus starts oscillating
is then given by \bea
T_{osc}^{X_N}&=&\left( \frac{90}{4\pi^2}  \right)^{1/4} g_\ast^{-1/4}(T^{X_N}_{osc}), \nonumber \\
&=&2.70 \times 10^{12} \left( \frac{228.75}{g_\ast(T^{X_N}_{osc})}
\right)^{1/4} \left(  \frac{m_{X_N}}{600 \, m_{3/2}} \right)^{1/2}
\; \mathrm{GeV}. \label{toscxn} \eea From this we find the entropy
density \bea
s(T_{osc}^{X_N})&=&\frac{2 \pi^2}{45} g_{osc} T_{osc}^3, \\
&=& 1.98 \times 10^{39} \left( \frac{g_{osc}}{228.75}
\right)^{1/4}  \left( \frac{m_{X_N} }{600 \, m_{3/2}}
\right)^{3/2} \; \mathrm{GeV}^3, \eea and the comoving abundance
is then \bea
Y^{(0)}_{X_N}&=&= {\half m_{X_N} f_{X_N}^2}{s^{-1}(T_{osc}^{X_N})}, \nonumber \\
&=&4.51 \times   10^4 \left( \frac{228.75}{g_{\ast}( T_{osc} )}
\right)^{1/4} \left( \frac{ f_{X_N} }{ m_p }  \right)^2 \left(
\frac{ 600 \, m_{3/2} }{ m_{X_N} } \right)^{1/2}, \eea

The oscillating modulus will quickly come to dominate the
radiation density and the temperature at this time is given by \be
T^{X_N}_{eq}=1.80 \times 10^{12}  \left(
\frac{228.75}{g_\ast(T^{X_N}_{osc})} \right)^{1/4} \left(
\frac{m_{X_N}}{600 \, m_{3/2}} \right)^{1/2} \left(
\frac{f_{X_N}}{m_p} \right)^2 \; \mathrm{GeV}, \ee so that we see
once the modulus starts coherent oscillations it quickly overtakes
the energy density (i.e., $T_{eq}^{X_N} \approx T_{osc}^{X_N}$).

\subsection{Meson and Light Moduli Oscillations}
Because the meson and light moduli are approximately degenerate in
mass (i.e. $m_\phi = m_{X_i}$) they will begin to oscillate at the
same time, \be 3 H^2 = 3 \left( \frac{2}{3} m_\phi \right) ^2 =
m_p^{-2} \left( \frac{\pi^2}{30} g_\ast(T^\phi_{osc})
\left(T_{osc}^{\phi}\right)^4 +m_{X_N} Y_{X_N} s(T^\phi_{osc})
\right). \ee Noting that the radiation term has already become
negligible compared to the heavy modulus density we find the
temperature at this time is given by {\footnotesize \bea
T^\phi_{osc}&=&\left( \frac{30}{\pi^2} \right)^{1/3}  \left[ \frac{ m_\phi^2 m_p^2}{g_{\ast s}(T^\phi_{osc}) m_{X_N} Y_{X_N}}  \right]^{1/3},\\
&=& 8.24 \times 10^{10} \left(
\frac{228.75}{g_\ast(T^{X_N}_{osc})} \right)^{1/4} \left(
\frac{m_{\phi}}{ 2 \, m_{3/2}} \right)^{2/3} \left( \frac{600 \,
m_{3/2}}{m_{X_N}} \right)^{1/6}  \left(
\frac{m_p}{f_{X_N}}\right)^{2/3}  \; \mathrm{GeV} \eea } which is
in excellent agreement with the exact answer obtained numerically
(including radiation) $T^\phi_{osc}=9.97 \times 10^{10}$. The
entropy density at this time is \be s(T_{osc}^{\phi})=5.62 \times
10^{34} \left( \frac{g_\ast(T^{X_N}_{osc})}{228.75} \right)^{1/4}
\left( \frac{m_{\phi}}{2 \, m_{3/2}} \right)^{2} \left(  \frac{600
\, m_{3/2} }{m_{X_N}} \right)^{1/2} \left( \frac{m_p}{f_{X_N}}
\right)^{2}  \; \mathrm{GeV}^3, \ee The meson $\phi$ initial
abundance is then \be Y^{(0)}_{\phi}=5.30  \times 10^6 \left(
\frac{228.75}{g_{\ast}( T_{osc} )} \right)^{1/4} \left( \frac{
f_{\phi} }{ m_p }  \right)^2 \left( \frac{ f_{X_N} }{ m_p }
\right)^2  \left( \frac{ m_{X_N} }{ 600 \, m_{3/2} } \right)^{1/2}
\left( \frac{ 2 \, m_{3/2} }{ m_{\phi} } \right). \ee The light
moduli will begin coherent oscillations at roughly the same time
as the meson.  Their abundance is then given by {\footnotesize
\bea
Y^{(0)}_{X_i}&=&(N-1) \, Y^{(0)}_{\phi} \nonumber \\
&=& 5.25  \times 10^8 \left( \frac{N-1}{99} \right) \left(
\frac{228.75}{g_{\ast}( T_{osc} )} \right)^{1/4} \left( \frac{
f_{X_i} }{ m_p }  \right)^2 \left( \frac{ f_{X_N} }{ m_p }
\right)^2  \left( \frac{ m_{X_N} }{ 600 \, m_{3/2} } \right)^{1/2}
\left( \frac{ 2 \, m_{3/2} }{ m_{X_i} } \right), \eea } where we
have implicitly assumed that because the masses of the meson and
light moduli are approximately degenerate they will have equal oscillation
amplitudes\footnote{We note that initially this may not be the
case, but at the onset of coherent oscillations (much less than a
Hubble time) the system will settle into this symmetric
configuration.}.

\subsection{Heavy Modulus Decay}
Once the Hubble parameter decreases to the point when $H \approx
\Gamma_{X_N}$, the heavy modulus decays and from  (\ref{treqn})
the corresponding reheat temperature is, {\footnotesize \be
T_r^{X_N}=41.40 \;   \left(  \frac{10.75}{g_{\ast }(T_r^{X_N})}
\right)^{1/4} \left( \frac{\cd_{X_N}}{1.6}  \right)^{1/2}  \left(
\frac{ m_{X_N} }{ 600 \; m_{3/2} }  \right)^{3/2} \left(
\frac{m_p}{f_\phi} \right)^{1/2} \left( \frac{100}{N}
\right)^{1/4} \; \mathrm{GeV}. \ee } To understand the $N$ and
$\phi$ dependence in this expression, we note that from
(\ref{treqn}) the reheat temperature includes the factor, \be
\left(\frac{m_{X_N} Y_{X_N} + m_{\phi} Y_\phi + m_{X_i} Y_{X_i}
}{m_{X_N} Y_{X_N} }\right)^{1/4} \ee Using that the meson and
light moduli have degenerate mass and therefore equal oscillation
amplitudes (i.e. $m_{X_i} Y_{X_i} = (N-1) m_\phi Y_\phi$) we find
\be \label{approxfrac} \left( 1+N \frac{ m_\phi Y_\phi }{ m_{X_N}
Y_{X_N} }  \right)^{-1/4} \approx \left( N \frac{ m_\phi Y_\phi }{
m_{X_N} Y_{X_N} } \right)^{-1/4} \ee which leads to the parametric
dependence in the reheat temperature.

Using (\ref{bigdel}) the entropy increase resulting from the heavy
modulus decay is \be \bsl \Delta_{X_N}=4.35  \times 10^{10}
\left( \frac{g_{\ast}( T_{r}^{X_N} )}{10.75} \right)^{1/4}  \left(
\frac{228.75}{g_{\ast}( T_{osc}^{X_N} )} \right)^{1/4} \left(
\frac{ f_{X_N} }{ m_p }  \right)^2 \left( \frac{1.6}{\cd_{X_N}}
\right)^{1/2} \\ \times \left( \frac{ 600 \; m_{3/2} } { m_{X_N} }
\right) \left(  \frac{ f_\phi } {m_p}  \right)^{1/2}  \left(
\frac{ N } {100}  \right)^{1/4},
\end{split}
\ee where we have again used (\ref{approxfrac}). Therefore, after
the decay the other moduli abundances are given by {\footnotesize
\bea
Y^{({X_N})}_{\phi}&=&\Delta_{X_N}^{-1} Y^{(0)}_{\phi}, \nonumber \\
&=& 1.22 \times 10^{-4} \left( \frac{10.75}{g_{\ast}( T_{r}^{X_N}
)} \right)^{1/4} \left( \frac{\cd_{X_N}}{1.6} \right)^{1/2}
\left( \frac{ f_{\phi} }{ m_p }  \right)^{3/2} \left(
\frac{m_{X_N}}{600 \, m_{3/2} }  \right)^{3/2} \left(  \frac{2 \,
m_{3/2} }{m_\phi}  \right) \left(  \frac{100}{N}  \right)^{1/4}
\;\;\;\;\;\;\;\; \eea } {\footnotesize \bea
Y^{({X_N})}_{X_i}&=&\Delta_{X_N}^{-1} Y^{(0)}_{X_i}, \nonumber \\
&=& 1.21 \times 10^{-2}  \left( \frac{10.75}{g_{\ast}( T_{r}^{X_N}
)} \right)^{1/4} \left( \frac{\cd_{X_N}}{1.6} \right)^{1/2} \left(
\frac{ f_{X_i} }{ m_p }  \right)^{3/2} \left(  \frac{m_{X_N}}{600
\, m_{3/2} }  \right)^{3/2} \left(  \frac{2 \, m_{3/2} }{m_{X_i}}
\right) \left(  \frac{N}{100}  \right)^{3/4}, \;\;\;\;\;\;\;\;
\eea }where we have again used $N-1 \approx N$. There is also a
decay to gravitinos with branching ratio
$B_{3/2}^{(X_N)}=0.2\%=0.002$. The corresponding comoving
abundance is thus, {\footnotesize \bea
Y_{3/2}^{(X_N)}&=&2 \times B_{3/2} \times \frac{Y_{X_N}^{(0)}}{\Delta_{X_N}}, \nonumber \\
&=&1.45 \times 10^{-9} \left( \frac{B_{3/2}}{0.07 \%} \right)
\left( \frac{10.75}{g_{\ast}( T_{r}^{X_N} )} \right)^{1/4} \left(
\frac{ \cd_{X_N} }{ 1.6 }  \right)^{1/2} \left( \frac{m_{X_N}
}{600 \, m_{3/2} } \right)^{1/2}
 \left( \frac{ m_p }{ f_\phi} \right)^{1/4} \left( \frac{ 100 }{N } \right)^{1/4},
\eea }

\subsection{Meson Decay}
When the meson decays, its contribution to the total energy
density will be less than that of the other $N-1$ light moduli.
The universe will be matter dominated before and after the decay,
but because the two energy sources are comparable there is a
somewhat significant entropy production. The meson decay reheats
the universe to a temperature \bea T_r^\phi &=&134  \times \left(
\frac{100} {N}\right)^{1/4} \left( \frac{10.75}{g_\ast(T_r)}
\right)^{1/4} \left( \frac{\cd_\phi}{711.6} \right)^{1/2} \left(
\frac{m_{\phi}}{2 \, m_{3/2}} \right)^{3/2}  \; \mathrm{MeV}. \eea
The entropy increase is given by \be \Delta_\phi=121 \times \left(
\frac{\cd_{X_N}}{1.6}  \right)^{1/2}  \left(
\frac{711.6}{\cd_{\phi}}  \right)^{1/2}  \left( \frac{m_{X_N}}{600
\, m_{3/2}}  \right)^{3/2} \left(  \frac{2 \, m_{3/2} }{m_\phi}
\right)^{3/4}
 \left(  \frac{f_\phi}{m_p}  \right)^{3/2}.
\ee The decay of the meson will further dilute the other moduli,
we find \bea
Y_{X_i}^{(\phi)} &=& \Delta_{X_N}^{-1} \Delta_{\phi}^{-1} Y_{X_i}^{(0)}, \nonumber \\
&=& 9.94 \times 10^{-5}  \left(  \frac{N}{100}  \right)^{3/4}
\left( \frac{10.75}{g_\ast(T_r)} \right)^{1/4} \left(
\frac{\cd_\phi}{711.6} \right)^{1/2} \left( \frac{ 2 \,
m_{3/2}}{m_{X_i}} \right)^{1/4}. \eea The decay of both the meson
and the light moduli to gravitinos is kinematically suppressed, so
that the only source of gravitinos comes from the decay of the
heavy modulus. This abundance after the decay of the meson is then
{\footnotesize \bea Y_{m_{3/2}}^{(\phi)}&=&\Delta_\phi^{-1}
Y_{m_{3/2}}^{(X_N)}, \nonumber \\  &=& {1.19 \times 10^{-11} }
\left( \frac{B^{(X_N)}_{3/2}}{ 0.07\%} \right)  \left(
\frac{100}{N}  \right)^{1/4}  \left( \frac{10.75}{g_{\ast}( T_{r}
)} \right)^{1/4} \left( \frac{\cd_{\phi}}{711.6} \right)^{1/2}
\left( \frac{ m_{\phi} }{ 2 \, m_{3/2} } \right)^{3/4}  \left(
\frac{600 \, m_{3/2}}{m_{X_N}}  \right) \left( \frac{m_p}{f_\phi}
\right)^{7/4}.\;\;\;\;\;\;\; \eea }

\subsection{Light Moduli Decays}
The decay of the light moduli results in a reheating temperature
\be \label{lightreheat} T_r^{X_i}=31.7 \times \left( \frac{ 10.75
}{ g_\ast(T_r)  } \right)^{1/4} \left( \frac{ m_{X_i}  }{ 2 \,
m_{3/2}  } \right)^{3/2} \left( \frac{ \cd_{X_i}  }{ 4  }
\right)^{1/2} \; \mathrm{MeV}, \ee which agrees with the bounds
set by BBN (i.e. $T_r^{X_i} > 1 \, \mathrm{MeV}$). The resulting
entropy production is \be \Delta_{X_i}=417.7  \times \left(
\frac{\cd_\phi}{711.6  } \right)^{1/2} \left( \frac{ 4  }{
\cd_{X_i}  } \right)^{1/2} \left(  \frac{2 \, m_{3/2}}{m_{X_i}}
\right)^{3/4} \left( \frac{N}{100}  \right)^{3/4}. \ee The new
gravitino abundance is given by {\footnotesize \bea
Y^{(X_i)}_{m_{3/2}}&=&\Delta_{X_i}^{-1} Y^{(\phi)}_{m_{3/2}}, \\
&=& 2.86 \times 10^{-14}  \left(  \frac{B_{3/2}^{(X_N)} }{0.07 \%
} \right) \left(  \frac{ 100 }{N } \right)  \left(  \frac{ 10.75
}{ g_\ast(T_r)   } \right)^{1/4} \left(  \frac{ m_\phi }{ 2 \,
m_{3/2}   } \right)^{3/2} \left( \frac{600 \, m_{3/2}}{m_{X_N}}
\right)  \left( \frac{\cd_{X_i}}{4} \right)^{1/2} \left(
\frac{m_p}{f_\phi} \right)^{7/4} \eea } which is small enough to
avoid the gravitino problem. The light moduli will decay into LSPs
yielding an abundance \bea
Y_{LSP}^{(X^{i})}&=&\Delta_{X_i}^{-1} B_{LSP}^{X_i}  Y_{X_i}^{(\phi)} , \nonumber \\
&=& 1.19 \times 10^{-7}  \left( \frac{B^{X_i}_{LSP}}{25 \%}
\right) \left( \frac{10.75}{g_\ast(T_r)} \right)^{1/4} \left(
\frac{m_{X_i}}{2 \, m_{3/2}} \right)^{1/2} \left(
\frac{\cd_{X_i}}{4} \right)^{1/2}, \label{lspabundance} \eea where
$B_{LSP}^{X_i}$ is the branching ratio for the decay of the light
moduli to LSPs. This corresponds to a number density at the time
of decay of $n_{LSP}=1.79 \times 10^{-11} \, \mathrm{GeV}^3$.

As we noted in the text, this abundance is produced below the
freeze-out temperature of the LSPs (non-thermal production) and is
greater than the critical density (\ref{nann}) for annihilations
to take place, which is $n_{X_i}^c=4.12 \times 10^{-15} \,
\mathrm{GeV}^3$. Thus, the LSPs will quickly annihilate (in less
than a Hubble time) and the final abundance will be given by the
critical value.

Thus, the relic density coming from the decay of the light moduli
is given by {\footnotesize \bea
\Omega_{LSP}&=& \frac{ {m_{LSP} Y^c_{LSP} s_0}}{\rho_c}, \nonumber \\
&=& 0.26 \, h^{-2} \left( \frac{m_{LSP}}{100 \, \mathrm{GeV}}
\right)^3 \left( \frac{10.75}{g_\ast(T_r)} \right)^{1/4}  \left(
\frac{3.26 \times 10^{-3} {\rm GeV}^{-2}}{\sigma_0} \right) \left(
\frac{4}{\cd_{X_i}} \right)^{1/2}  \left( \frac{2 \,
m_{3/2}}{m_{X_i}} \right)^{3/2} \left(  \frac{100 \,
\mathrm{TeV}}{m_{3/2} }\right)^{3/2}, \eea } where $s_0$ and
$\rho_{c}$ are the entropy density and critical density today
respectively, and we have used the experimental value $\rho_c /
s_0= 3.6 \times 10^{-9} h^2 \; \mathrm{GeV}$ where $h$
parametrizes the Hubble parameter today with median value
$h=0.71$.

\section{Couplings and Decay Widths of the Moduli and Meson Fields\label{appendixa}}

In this section, we discuss the moduli couplings to MSSM particles
and then calculate their decay widths in terms of the microscopic
parameters of the $G_2$-MSSM framework. This will motivate the
benchmark values used for numerical results throughout the paper.
We will find that the moduli decay into scalars is very important.

\subsection{Moduli Couplings}
Let us first consider the couplings associated with $N$
eigenstates $X_j$ of the geometric moduli $s_i$. For simplicity,
we neglect the small mixing with the meson modulus $\phi$ (we will
return to that later). First consider the moduli coupling to gauge
bosons through the gauge kinetic function $f^{sm}$. The relevant
term is:
\begin{eqnarray}
 {\cal L}&\supset & -\frac{1}{4}{\rm Im}(f_{sm}) F^a_{\mu\nu}F^{a\mu\nu}\\
   &=&-\frac{1}{4}\langle {\rm Im}(f_{sm})\rangle
   F^a_{\mu\nu}F^{a\mu\nu}-\frac{1}{4}\sum_{i}N_i^{sm} \delta s_i
   F^a_{\mu\nu}F^{a\mu\nu}
\end{eqnarray}
where we have expanded the moduli as $s_i=\langle s_i\rangle
+\delta s_i$. After normalizing the gauge fields and the moduli
fields, the interaction term can be written as:
\begin{eqnarray}
   {\cal L}_{X_j gg}&=&\frac{1}{4\,f_{sm}}\sum_{i=1}^{N}
N_i^{sm}\sqrt{\frac{2 \langle s_i\rangle}{3a_i}}U_{ij}\;X_j F^a_{\mu\nu}F^{a\mu\nu}\\
     &=& \frac{\sqrt{7}}{6\sqrt{2}}{\cal B}\;{\cal C}_j\;X_j
     F^a_{\mu\nu}F^{a\mu\nu},
\end{eqnarray}
where $\cal B$ and ${\cal C}_j$ are defined as:
\begin{eqnarray}
{\cal B}&\equiv& \left(\sum_{i=1}^{N}
\frac{N_i^{sm}}{N_i}\;a_i\right)^{-1}\\
{\cal C}_j&\equiv&\sum_{i=1}^{N} \frac{N_i^{sm}}{N_i} \;(\vec
X_N)_i (\vec X_j)_i .
\end{eqnarray}
For the heavy modulus, since $(X_N)_i^2=\frac{3}{7}a_i$, we have
${\cal C}_N=\frac{3}{7}{\cal B}^{-1}$ while for the light moduli
$X_i, i=1,\cdots,(N-1)$, it is easy to show:
\begin{equation}\label{eq:sumC}
\sum_{i=1}^{N-1}{\cal C}_i^2=l^2\sin^2\theta,
\end{equation}
where $l$ is the length of the vector $\vec X_N^{\;'}$ defined as
$(\vec X_N^{\;'})_i\equiv(\vec X_N)_i\,N_i^{sm}/N_i$ and $\theta$
is the angle between $\vec X_N^{\;'}$ and $\vec X_N$. So,
generically ${\cal C}_i$ are less than one. There are two extreme
cases: one when $N_i^{sm}=k N_i $ in which the moduli couplings to
gauge bosons vanish since the vector $\vec X_N$ is orthogonal to
$\vec X_j$, and the other when $\vec X_N^{\;'}$ equal to one of
the $X_i$'s in which all ${\cal C}_i$'s are zero except one.

For the couplings to gauginos, the dominant contribution comes
from the following terms in the lagrangian:
\begin{eqnarray}
  {\cal L} \supset -\frac{i}{4}\partial_i f_{sm} F^i \lambda^a
  \lambda^a + h.c.
\end{eqnarray}
where $\partial_i f_{sm}=N^{sm}_i$ and $-i$ arises because of the
convention of the moduli chiral fields $z_i=t_i+is_i$ we used.
Expanding the $F$-terms of the moduli fields around their
\emph{vevs}, we have:
\begin{eqnarray}
   F^i &=& \langle F^i\rangle + \langle \partial_{s_k} F^i \rangle
   \delta s_k
\end{eqnarray}
The derivative of the $F$-term can be calculated as follows:
\begin{eqnarray}
  \partial_{s_k} F^i &=& \partial_{s_k} \Big( e^{K/2} K^{i\bar j} (K_{\bar
  j}W^* + W_{\bar j}^*)\Big)\nonumber\\
  &= & -ie^{-i\gamma}m_{3/2} \;\left(\frac{4}{3}\frac{a_i}{N_i} N_k \nu^2 \frac{\tilde z}{\tilde x} +\frac{4}{3}\frac{N_k}{N_i}(3a_i-2\delta_{ik})\nu\frac{\tilde y}{\tilde x}+\frac{N_k}{N_i}(3a_i-2\delta_{ik})\right)\nonumber\\
  &=& -ie^{-i\gamma}m_{3/2} \;\left(-\frac{4}{3}s_i N_k b_1b_2\nu -3\frac{N_k}{N_i}a_i + 2\delta_{ik}+\cdots\right)
\end{eqnarray}
where in the last line, the subleading terms are not explicitly
shown. $\gamma$ is the phase in the superpotential which will be
set to zero for simplicity without affecting any result here. We
have used the following equations:
\begin{eqnarray}
  &&\partial_{s_i} K = -\frac{3a_i}{s_i}, \quad K^{i\bar j}= \frac{4 s_i^2}{3 a_i}\delta_{i\bar j}, \quad
  \partial_{s_k} K^{i\bar j}=\frac{2}{s_k}
  K^{i\bar j}\delta_{ik}
\end{eqnarray}
After normalizing the moduli fields and the gauge fields, the
couplings are given by:
\begin{eqnarray}
  {\cal L}_{X_i \lambda\lambda} &\approx&\frac{1}{4}\sqrt{\frac{2}{3}}m_{3/2}
  \left[\left(\frac{4}{3}\nu^2 b_1b_2 \right)\sum_{k=1}^{N}a_k^{1/2}U_{ki}-\frac{1}{f_{\rm sm}} 2\nu
  \sum_{k=1}^{N} \frac{N_k^{sm}}{N_k}a_k^{1/2}U_{ki}
  \right]X_i
  \lambda^{a}\lambda^{a}+h.c.\nonumber\\
  &=&\frac{\sqrt{14}}{12}m_{3/2}\left[\frac{4}{3}\nu^2 b_1b_2(\vec X_N\cdot\vec X_i)-2
  {\cal B}\,(\vec X_N'\cdot \vec X_i)
  \right]X_i
  \lambda^{a}\lambda^{a}+h.c.
\end{eqnarray}
For the light moduli fields, the first term vanishes and the
couplings turn out to be:
\begin{eqnarray}
  {\cal L}_{X_l\lambda\lambda} \approx -\frac{\sqrt{14}}{6} {\cal B}\;{\cal
  C}_i m_{3/2}\;X_i
  \lambda^{a}\lambda^{a}+h.c.
\end{eqnarray}
For the heavy modulus field, the first dot product is unity and
the coupling is:
\begin{eqnarray}
    {\cal L}_{X_l\lambda\lambda}\approx\frac{\sqrt{14}}{12}m_{3/2}\left(\frac{4}{3}\nu^2 b_1b_2\right)\;X_i
  \lambda^{a}\lambda^{a}+h.c.
\end{eqnarray}

The moduli couplings to other MSSM particles can be derived
generically by expanding all the moduli around their \emph{vevs}
in the supergravity lagrangian:
\begin{eqnarray}
  {\cal L}&\supset& \tilde K_{\bar\alpha\beta} {\cal D}_{\mu}{\tilde f}^{*\bar\alpha} {\cal D}^{\mu}\tilde f^{\beta}
+i\tilde K_{\bar\alpha\beta} f^{\dagger\bar\alpha}{\bar\sigma}^{\mu}{\cal D}_{\mu} f^{\beta}-V(\tilde f^*,\tilde f)+\cdots
\end{eqnarray}
where $f^{\alpha}$ and $\tilde f^{\alpha}$ are fermions and their superpartners. The other derivative terms involving moduli and matter fields
are not explicitly shown for simplicity. The relevent coupling here are the moduli-sfermion-sfermion
coupling and the moduli-fermion-fermion coupling. They are found to be
%where $m_p=2.44 \times 10^{18} \; GeV$ is the reduced Planck mass and we have expanded the K\"{a}hler potential around the moduli
%part $K_0$. In general the matter Kahler metric depends on both
%the geometric moduli $s_i$ and the hidden sector moduli $\phi$.
%Then we can derive the geometric moduli couplings to matter
%chiral fields as follows
\begin{eqnarray}
 {\cal L} &\supset& \partial_{s_i} \tilde K_{\bar\alpha\beta}\left[\delta s_i\;\partial_{\mu}{\tilde f}^{*\bar\alpha}\partial^{\mu}{\tilde
f}^{\beta}+i\delta s_i\;f^{\dagger\bar\alpha}{\bar\sigma}^{\mu}\partial_{\mu}f^{\beta}\right]-\partial_{s_i} {m'}^2_{\bar\alpha\beta}\,\delta s_i{\tilde f}^{*\bar\alpha} {\tilde f}^{\beta}+\cdots\\
&=& {g'}_{X_i\tilde f\tilde f}^{\alpha}
\left[\partial_{\mu}(X_i\;{\tilde
f}^{*\bar\alpha}_c)\partial^{\mu}{\tilde f}^{\alpha}_c+c.c.+ i
X_i\;{\bar
f}^{\bar\alpha}_c{\bar\sigma}^{\mu}\partial_{\mu}f^{\alpha}_c\right]-
g_{X_i\tilde f\tilde f}^{\alpha}X_i {\tilde f}^{*\bar\alpha}_c
{\tilde f}^{\alpha}_c+\cdots\label{coup-geom-moduli3}
\end{eqnarray}
where $\tilde f^{\alpha}_c$ and $f^{\alpha}_c$ are the canonical
normalized fields. For simplicity, we consider the Kahler metric
to be diagonal $\tilde K_{\bar\alpha\beta}=\tilde
K_{\alpha}\delta_{\bar\alpha\beta}$, then
\begin{eqnarray}
  g_{X_j\tilde f\tilde f}^{\alpha}&\approx& m_{3/2}^2\partial_{s_i}\log({\tilde K}_{\alpha})\sqrt{\frac{2s_i^2}{3a_i}}U_{ij}\nonumber\\
  &=&\frac{\sqrt{14}}{3}m_{3/2}^2\,(\vec X_N'')^{\alpha}\cdot \vec X_j \\%\sum_{i=1}^{N}\frac{\xi_{i,\alpha}}{a_i}(X_N)_i(X_j)_i
  {g'}_{X_j\tilde f \tilde f}^{\alpha}&=&\frac{\sqrt{14}}{6}\,(\vec X_N'')^{\alpha}\cdot \vec X_j
\end{eqnarray}
where $(\vec X_N'')^{\alpha}_i\equiv\xi_{i,\alpha}(X_N)_i/a_i$ and
$\xi_{i,\alpha}\equiv s_i\partial_{s_i}\log(K_{\alpha})$. In this
calculation, we have used the fact that $\partial_{\phi_0}{\tilde
K}_{\alpha}=0$ and have neglected terms involving $F$-terms of
geometric moduli $F^i$ which are suppressed relative to $m_{3/2}$.

For the couplings to the higgs doublets, there are differences
from other scalars. The kinetic terms and the mass terms for the
higgs fields in the MSSM can be written as:
\begin{eqnarray}
 {\cal L} &\supset& {\tilde K}_{H_u}\left[\partial_{\mu}H_u^{*}\partial^{\mu}
H_u+i{\bar {\tilde H}}_u{\bar\sigma}^{\mu}\partial_{\mu} {\tilde H}_u\right]+\cdots\nonumber\\
&-&({\tilde K}_{H_d}^{-1}|\mu '|^2+{m'}_{H_u}^2)H_u^*H_u+(H_u\leftrightarrow H_d)\nonumber\\
&-&(B\mu'H_d H_u+c.c.)
\end{eqnarray}
where
\begin{equation}
\mu '=m_{3/2}Z-{\bar F}^{\bar m} \partial_{\bar m}Z
\end{equation}
is only generated by the higgs bilinear term in the Kahler
potential\cite{Acharya:2008zi}. To derive the modular couplings to
higgs doublets, one needs $\partial_{s_i}\mu'$, which is:
\begin{eqnarray}
\partial_{s_i}\mu'=(\partial_{s_i}m_{3/2})Z+m_{3/2} \partial_{s_i}Z-(\partial_{s_i}{\bar F}^{\bar m})\partial_{\bar m}Z-{\bar F}^{\bar m} \partial_{s_i}\partial_{\bar m}Z
\end{eqnarray}
One can see that the second and the third terms are of order
$m_{3/2}$ while the rest are suppressed. Therefore, the dominant
contribution is:
\begin{eqnarray}
\partial_{s_i}\mu'&\approx& \frac{1}{2}m_{3/2}(\partial_{s_m}Z) \;\left(-\frac{4}{3}s_m N_i b_1b_2\nu + 4\delta_{im}\right)
\end{eqnarray}
For simplicity, taking all the phases of the superpotential and
that of $Z$ to be vanishing, we find:
\begin{eqnarray}
-{\cal L}&\supset&
%-{g'}_{X_i H_u H_u}\left[\partial_{\mu}(X_i H_u^{*})\partial^{\mu}
%H_u + c.c. +iX_i{\bar {\tilde H}}_u{\bar\sigma}^{\mu}\partial_{\mu} {\tilde H}_u\right] +
g_{X_j H_u H_u} X_j H_u^* H_u \\
g_{X_j H_u H_u}&\approx& m_{3/2}^2\bigg[Z_{\rm eff}^2\partial_{s_m} \log Z\left(-\frac{4}{3}s_m N_i b_1b_2\nu + 4\delta_{im}\right)-Z_{\rm eff}^2\partial_{s_i}\log{\tilde K}_{H_d}\nonumber\\
&+&\partial_{s_i}\log{\tilde K}_{H_u}\bigg]\sqrt{\frac{2s_i^2}{3a_i}}U_{ij}\nonumber\\
&=& \frac{\sqrt{14}}{3}m_{3/2}^2 Z_{\rm eff}^2\bigg(-\frac{4}{3}\nu^2 b_1 b_2\left(\sum_{m=1}^N \zeta_m\right)\vec X_N\cdot \vec X_j + 4 \vec X_N'''\cdot \vec X_j\nonumber\\
&-&(\vec X_N'')^{H_d}\cdot \vec X_j\bigg)+ \frac{\sqrt{14}}{3}m_{3/2}^2 (\vec X_N'')^{H_u} \cdot \vec X_j
\end{eqnarray}
where $(\vec X_N''')_i\equiv\frac{\zeta_i}{a_i}(X_N)_i$ and
$\zeta_i\equiv s_i\partial_{s_i}\log(Z)$. We also use the fact
that $\partial_{\phi_0}Z=0$ and the $F$-terms $F_i/m_p \ll
m_{3/2}$ for geometric moduli. To get the corresponding couplings
for $H_d$, we can simply replace $H_u$ by $H_d$ in the above
equations. The coupling of moduli to higgs through the kinetic
term is similar to the non-higgs scalar
\begin{eqnarray}
{g'}_{X_i H_u H_u}^{\alpha}&=&\frac{\sqrt{14}}{6}\,(\vec X_N'')^{H_u}\cdot \vec X_i
\end{eqnarray}

Let us now consider the $B\mu$ term, which is given by:
\begin{eqnarray}
B\mu'&=&(2m_{3/2}^2+V_0)Z-m_{3/2}{\bar F}^{\bar m} \partial_{\bar m} Z+m_{3/2} F^{m}[\partial_{m}Z-\partial_{m}\log({\tilde K}_{H_u}{\tilde K}_{H_d})\,Z]\nonumber\\
&-&{\bar F}^{\bar m}F^{n}[\partial_{\bar m}\partial_{n}Z-\partial_{n}\log({\tilde K}_{H_u}{\tilde K}_{H_d})\partial_{\bar m}Z].
\end{eqnarray}
The corresponding derivative is given by:
\begin{equation}
\partial_{s_i}B\mu'\approx \frac{1}{2} m_{3/2}^2 \,Z\,\partial_{s_i}\log({\tilde K}_{H_u}{\tilde K}_{H_d})\left(-\frac{4}{3}s_i N_k b_1b_2\nu + 2\delta_{ik}\right) + 2m_{3/2}^2\partial_{s_i}Z,
\end{equation}
which gives rise to the coupling:
\begin{eqnarray}
-{\cal L}&\supset& g_{X_j H_d H_u} X_j H_d H_u + c.c.\\
g_{X_j H_d H_u}&\approx & \frac{\sqrt{14}}{6}m_{3/2}^2\,Z_{\rm eff}\bigg(-\frac{4}{3}\nu^2 b_1b_2\left(\sum_{m=1}^N \xi^{H_u}_m\right)\vec X_N\cdot \vec X_j \nonumber\\
&+& 2(\vec X_N'')^{H_u}\cdot \vec X_j+(H_u\rightarrow H_d)+4 \vec X_N'''\cdot \vec X_j\bigg)
\end{eqnarray}
Besides the term mentioned above there is another coupling from
the bilinear term in the k\"{a}hler potential $K\sim Z(s_i) H_d
H_u+ h.c. $ \cite{Moroi:1999zb}. This term leads to a coupling:
\begin{eqnarray}
{\cal L}&\supset& g'_{X_j H_d H_u}
\partial_{\mu}X_j \partial^{\mu}(H_d H_u) +
  c.c.\label{coup-geom-moduli4}\\
g'_{X_j H_d H_u}%&=&\frac{Z}{K_{H_u}^{1/2}K_{H_d}^{1/2}}\partial_{s_i}\log(Z)\sqrt{\frac{2s_i^2}{3a_i}}U_{ij}\nonumber\\
  &=&\frac{\sqrt{14}}{6}Z_{\rm eff}\,\vec X_N'''\cdot \vec X_j
\end{eqnarray}
This coupling could be very important since it is proportional to
the moduli mass squared if equations of motion of $X_i$ are used.
Again for the coupling to be unsuppressed, the bilinear
coefficient $Z$ should have a sizable dependence on the geometric
moduli $s_i$, which is natural. This coupling is essential for
electroweak symmetry breaking in the $G_2$-MSSM.

\subsection{Meson Couplings}

In the $G_2$-MSSM framework, the hidden sector is not sequestered
from the visible sector and there are couplings between the hidden
sector meson field $\phi$ and various MSSM particles, which we
want to compute. First since the tree level gauge kinetic function
does not depend on $\phi$, there is no coupling to gauge bosons.
However there are couplings to the gauginos which depend on
$\partial_{\phi_0}$, which are computed to be
\begin{eqnarray}
  \partial_{\phi_0} F^{i}
  &=& -ie^{-i\gamma}\frac{4s_i}{3\phi_0}{\cal F}m_{3/2},\\
{\cal F}&=&\frac{2Q\P}{21P}+2+\frac{3}{P}+{\cal O}(\P^{-1}).
\end{eqnarray}
After normalization of fields, the coupling of meson to the
gauginos is given by:
\begin{eqnarray}
  {\cal L}_{\delta\phi_0\lambda\lambda}= e^{-i\gamma}\frac{1}{3\sqrt{2}\phi_0}{\cal F}m_{3/2}\delta \phi_0 \lambda\lambda
\end{eqnarray}
We now move on to the couplings of the meson field to scalars. We
will assume that the K\"{a}hler metric and the higgs bilinear $Z$
do not depend on $\phi_0$. We then have for the non-higgs scalars:
\ba {\cal L}_{\delta\phi_0\tilde f\tilde{f}} &=&
\frac{1}{\sqrt{2}\tilde{K}_{\alpha}}\,\frac{\partial
m^{'2}_{\alpha}}{\partial
\phi_0}\delta{\phi_0}\tilde{f}^*\tilde{f}\nonumber\\
&=& \sqrt{2}m_{3/2}\,(\partial_{\phi_0}\,m_{3/2})\delta{\phi_0}\tilde{f}^*\tilde{f}\nonumber\\
&\approx&
\sqrt{2}m_{3/2}^2\phi_0(1+\frac{2}{3\phi_0^2})\delta{\phi_0}\tilde{f}^*\tilde{f}\label{phi-f-f}\ea
In the above, we have neglected terms proportional to $F_i/m_p$
which are $\ll m_{3/2}$. There are various kinds of couplings of
the meson to the Higgs fields $H_u$ and $H_d$. The coupling
originating from the term $\int d^4\theta\,(ZH_uH_d+c.c)$ does
\emph{not} give rise to any contribution since $Z$ is assumed to
be independent of $\phi_0$. The couplings ${\cal
L}_{\delta\phi_0H_u^*H_u}$ and ${\cal L}_{\delta\phi_0H_d^*H_d}$
are computed as follows:
\ba {\cal L}_{\delta\phi_0H_u H_u} &=&g_{\delta\phi_0 H_u H_u}\delta{\phi_0}\tilde{H}_u^*\tilde{H}_u\nonumber\\
g_{\delta\phi_0 H_u H_u}&=&\frac{1}{\sqrt{2}\tilde{K}_{H_u}}\,\frac{\partial
({\tilde K}_{H_d}^{-1}|\mu'|^2+m^{'2}_{H_u})}{\partial
\phi_0}\nonumber\\
&\approx&
\sqrt{2}(Z_{\rm eff}^2+1)\,m_{3/2}^2\phi_0\left[(1+\frac{2}{3\phi_0^2})+(\frac{Z_{\rm eff}^2}{Z_{\rm eff}^2+1})\frac{2\cal
F}{3\phi_0^2}\sum_{i=1}^N\zeta_i\right]\label{phi-Hu-Hu}\ea ${\cal
L}_{\delta\phi_0\hat{H}_d^*\hat{H}_d}$ can be obtained from the
above by replacing $H_u$ with $H_d$. Again, we have neglected
terms proportional to $F_i/m_p$. Finally, we look at the coupling
${\cal L}_{\delta\phi_0H_dH_u}$. It is given by: \ba
{\cal L}_{\delta\phi_0H_dH_u} &=&g_{\delta\phi_0 H_d H_u}\delta{\phi_0}\tilde{H}_d\tilde{H}_u\nonumber\\
g_{\delta\phi_0 H_d H_u}&=&\frac{1}{\sqrt{2}(\tilde{K}_{H_u}\tilde{K}_{H_d})^{1/2}}\,\frac{\partial
(B\mu')}{\partial \phi_0}\nonumber\\
&\approx& \sqrt{2}m_{3/2}^2\,\phi_0 Z_{\rm eff}
\left[2(1+\frac{2}{3\phi_0^2})+\frac{\cal
F}{3\phi_0^2}\sum_{i=1}^N\,(\xi_i^{H_u}+\xi_i^{H_d})\right]\label{phi-Hd-Hu}\ea
The coupling ${\cal L}_{\delta\phi_0H_u^*H_d^*}$ can be computed
by taking the complex conjugate of the above expression.

\subsection{RG evolution of the couplings}

In the last subsection, we computed all the relevant couplings of
the moduli and meson at a high scale, presumably around the
unification scale. However, since the scale at which moduli decay
is much smaller than the unification scale, one should in
principle use the effective couplings at that scale to compute the
decay widths. The RG running of the moduli-scalar-scalar couplings
are especailly important for the third generation squarks and the
higgs doublets and are the main focus of this subsection. The
leading contribution to the $\beta$ functions are terms
proportional to $|y_t|^2$ and $g_3^2$\footnote{Here we have not
included the digrams proportional to $g_{X_j gg}$ and $g_{X_j
\tilde g\tilde g}$, since their contributions are relatively
smaller}, which are given below:
\begin{eqnarray}
  \beta(g_{X_j H_d H_u})&\approx& \frac{1}{16\pi^2}3|y_t|^2g_{X_j
  H_dH_u},\nonumber\\
  \beta(g'_{X_j H_d H_u})&\approx& \frac{1}{16\pi^2}3|y_t|^2g'_{X_j
  H_dH_u},\nonumber\\
  \beta(g_{X_j H_u H_u})&\approx&
  \frac{1}{16\pi^2}6|y_t|^2\left(g_{X_jH_u H_u}+X_t\right),\nonumber\\
  \beta(g'_{X_j H_u H_u})&\approx&
  \frac{1}{16\pi^2}6|y_t|^2 g'_{X_jH_u H_u},\nonumber\\
  \beta(g_{X_j \tilde Q_3 \tilde Q_3})&\approx& \frac{1}{16\pi^2}\bigg[g_{X_j
  \tilde Q_3\tilde Q_3}\left(2|y_t|^2-\frac{16}{3}g_3^2\right)+2|y_t|^2 X_t\bigg],\nonumber\\
  \beta(g'_{X_j \tilde Q_3 \tilde Q_3})&\approx& \frac{1}{16\pi^2}g'_{X_j
  \tilde Q_3\tilde Q_3}\left(2|y_t|^2-\frac{16}{3}g_3^2\right),\nonumber\\
\beta(g_{X_j \tilde u_3 \tilde u_3})&\approx&
\frac{1}{16\pi^2}\left[g_{X_j
  u_3u_3}\left(8|y_t|^2-\frac{16}{3}g_3^2\right)+4|y_t|^2 X_t\right],\nonumber\\
  \beta(g'_{X_j \tilde u_3 \tilde u_3})&\approx& \frac{1}{16\pi^2}g'_{X_j
  u_3u_3}\left(8|y_t|^2-\frac{16}{3}g_3^2\right),\label{beta-func}
\end{eqnarray}
where $X_t\equiv g_{X_jH_u H_u}+g_{X_j\tilde Q_3\tilde Q_3}+g_{X_j \tilde u_3 \tilde u_3}$.
For other beta functions not listed above, the RGE effects can be neglected.

To examine the RG effects on the moduli-scalar-scalar couplings,
we take all the weighted dot products involved in the
moduli-scalar-scalar couplings to be equal for
simplicity\footnote{The more general case will be studied later.},
\be {\vec X_{N}'''}\cdot \vec X_i=(\vec X_{N}'')^{\alpha}\cdot
\vec X_i=\Pi.\ee This is reasonable as their structure is very
similar. So the high scale couplings can be written as:
\begin{eqnarray}
  g_{X_jH_u H_u}&=&g_{X_jH_dH_d}=\frac{\sqrt{14}}{3}m_{3/2}^2 (3Z_{\rm
  eff}^2+1)\Pi\\
  g'_{X_jH_u H_u}&=&g'_{X_jH_dH_d}=\frac{\sqrt{14}}{6}\Pi\\
  g_{X_jH_d H_u}&=&\frac{4\sqrt{14}}{3}m_{3/2}^2 Z_{\rm
  eff}\Pi\\
  g'_{X_jH_d H_u}&=&\frac{\sqrt{14}}{6}Z_{\rm
  eff}\Pi
\end{eqnarray}
Using the beta functions given in Eq.(\ref{beta-func}), we can see
that at low scale $g_{X_j H_u H_u}$ is squashed because of the
large yukawa couplings. Similarly $g_{X_j \tilde Q_3\tilde Q_3}$
and $g_{X_j \tilde u_3 \tilde u_3}$ decrease significantly and
become negative at low scales.

One important thing to compute for moduli decay to light higgs is the effective coupling $g^{eff}_{X_j hh}$, which can be written in terms
of the couplings to higgs doublets
\begin{eqnarray}
g_{X_i hh}^{eff}&=& (g_{X_i H_u H_u}-2m_h^2 g'_{X_i H_u H_u})\cos^2\alpha+(g_{X_i H_d H_d}-2m_{h}^2g'_{X_i H_d H_d})
\sin^2\alpha\nonumber\\
&-&(g_{X_i H_d H_u}-m_{X_i}^2 {g'}_{X_i H_d H_u})\sin2\alpha\label{coup-xihh}
\end{eqnarray}
where all the couplings involved should be evaluated at low scales
and $\alpha$ is the higgs mixing angle. For the $G_2$-MSSM, the
higgs sector is almost in the ``decoupling region", which implies
$\alpha\approx \beta-\frac{\pi}{2}$. Now with universal boundary
condition for the weighted dot products for concreteness and
simplicity, the effective coupling of moduli to $hh$ final state
is given by:
\begin{eqnarray}
  g_{X_i hh}^{eff}&\approx& \frac{\sqrt{14}}{3}m_{3/2}^2 \left[(3Z_{\rm
  eff}^2+1)(\sin^2\alpha +K_1 \cos^2\alpha) - 2K_2 Z_{\rm
  eff}\sin(2\alpha)\right]\Pi
\end{eqnarray}
where $K_1$ and $K_2$ are the RG factors. To estimate these
factors, we take $y_t=1$, $\alpha_{unif}^{-1}=26.7$ and $Z_{\rm
eff}=1.58$, which is the same as the first Benchmark $G_2$-MSSM.
Then, typically we find $K_1\sim 0.2$ and $K_2\sim0.5$. For
readers not familiar with the details of the $G_2$-MSSM, it is
helpful to know that generically $\tan\beta\sim 1$ and $Z_{\rm
eff}\sim 1.5$. For the effective coupling to third generation
squarks, including the RG effects, we have:
\begin{eqnarray}
  g_{X_j \tilde u_3 \tilde u_3}^{eff}\approx g_{X_j \tilde Q_3 \tilde Q_3}^{eff} \sim
  \frac{\sqrt{14}}{3}m_{3/2}^2\Pi
\end{eqnarray}
where $g_{X_j \tilde f\tilde f}^{eff}\equiv g_{X_j \tilde f \tilde
f}-m_{\tilde f}^2 g'_{X_j \tilde f \tilde f}$. From the above RGE
results, we find that the couplings to the non-higgs scalars and
higgs should be roughly of the same order because of the large
radiative correction even when some of them are suppressed
relative to the other at the high scale boundary. Therefore, if
the couplings to scalars are large, then we should expect a
significant branching ratio of the moduli to LSPs.

For the coupling of the meson field to scalars, the $\beta$
functions are exactly the same. Similar to the analysis of light
moduli, we introduce factors $K_1$ and $K_2$ to account for the RG
effects on $g_{\phi H_u H_u}$ and $g_{\phi H_d H_u}$. Typically
one has $K_1\sim 0.25$ and $K_2\sim 0.5$. From
Eq.(\ref{phi-Hu-Hu}) and (\ref{phi-f-f}), we find the coupling
$g_{\phi H_u H_u}$ is at least $Z_{\rm eff}^2 {\cal F}\sim 30$
times larger than $g_{\phi \tilde f\tilde f}$ at the high scale.
Because of this large coupling $g_{\phi H_u H_u}$, even if the
couplings $g_{\phi \tilde Q_3 \tilde Q_3}$ and $g_{\phi \tilde u_3
\tilde u_3}$ are zero at the high scale, they can still be
generated at the low scale, which is proportional to $g_{\phi H_u
H_u}$ by a factor $K_3\sim 0.1$.

\subsection{Decay Rates of the Moduli}\label{sectiondecay}

Now that we have computed all the the relevant couplings for
moduli decay, we can proceed to compute the corresponding decay
widths. In the following, we give the result of decay widths for
all the moduli, calculated from the two-body width formulae. There
could be contribution from three-body decays, which is generally
small because of the phase space. Although certain three-body
decays, e.g. moduli to top quarks and
higgs\cite{Moroi:1999zb,Endo:2006qk} is relatively large, it is
still comparatively small in the current framework compared to the
two-body decay modes.

For light moduli $X_i, i=1,\cdots,(N-1)$, the total decay width
is
\begin{equation}
\Gamma(X_i)\equiv
\frac{D_{X_i}m_{X_i}^3}{m_p^2}=\frac{7}{72\pi}\left(N_G{\cal
A}_{1}^{X_i}+N_G{\cal A}_{2}^{X_i}+{\cal A}_{3}^{X_i}+{\cal
A}_{4}^{X_i}\right)\frac{m_{X_i}^3}{m_p^2},
\end{equation}
where ${\cal A}_{1}^{X_i}$, ${\cal A}_{2}^{X_i}$, ${\cal
A}_{3}^{X_i}$ and ${\cal A}_{4}^{X_i}$ are the corresponding
coefficients for the decays to gauge bosons $gg$, gauginos $\tilde
g\tilde g$, non-higgs scalars $\tilde f\tilde f$ and light higgs
bosons $hh$ respectively. They are given by:
\begin{eqnarray}
{\cal A}_{1}^{X_i}&=&\frac{1}{2}\left(\sum_{i=1}^{N}
\frac{N_i^{sm}}{N_i}\;a_i\right)^{-2}(\vec X_{N}'\cdot X_i)^2,\\
{\cal
A}_{2}^{X_i}&=&\left(\frac{m_{3/2}^2}{2m_{X_i}^2}\right)\left(\sum_{k=1}^{N}
\frac{N_k^{sm}}{N_k}\;a_k\right)^{-2}(\vec X_{N}'\cdot X_i)^2,\\
{\cal A}_{3}^{X_i}&\approx&\sum_{\alpha=\tilde t_L, \tilde t_R,\tilde b_L}3\left(\frac{m_{3/2}^4}{m_{X_i}^4}\right)
\left(1-4\frac{m_{\tilde f_{\alpha}}^2}{m_{X_i}^2}\right)^{1/2}\Pi^2,\\
{\cal A}_{4}^{X_i}&\approx&
\left(\frac{m_{3/2}^4}{2m_{X_i}^4}\right) \bigg[(3Z_{\rm
eff}^2+1)(\sin^2\alpha+K_1\cos^2\alpha)-2K_2 Z_{\rm
eff}\sin(2\alpha)\bigg]^2\Pi^2
\end{eqnarray}
Here, weighted dot products in the scalar couplings are assumed to
be equal and are denoted as $\Pi$ as in the last subsection. In
addition, the RGE effects on the couplings are included. In the
above result, the gaugino and gauge bosons are treated as
massless. The two-body decay to the standard model fermions is
suppressed by $(\frac{m_{f}}{m_{X_i}})^2 \ll 10^{-4}$, so it is
neglected in our result; even the top quark contribution is small.
For the decay to non-higgs scalars, naively there is a large
kinematic suppression since these scalars have mass close to
$m_{3/2}$. However, the RGE running significantly decreases the
third generation squark mass at the scale much lower than the
unification scale. In $G_2$ MSSM framework, the lightest stop is
$\tilde t_R$ which is about 4 times lighter than the gravitino.
It, therefore, has a large contribution to the partial width. In
addition, $\tilde Q_3$ ($\tilde b_L$ and $\tilde t_L$) are also
light enough such that they contribute to the decay width.

The above result for $\cal A$'s depend on the specific choices of
the fundamental parameters, such as $a_i$, $N_i$ and $N_i^{sm}$,
through several weighted dot products of vectors $\vec X_N$ and
$\vec X_i$. These quantities are different for different moduli.
However from Eq.(\ref{eq:sumC}) they are constrained by:
\begin{eqnarray}
\sum_{i=1}^{N} (\vec X_N'\cdot\vec X_i)^2 = |\vec X_N'|^2\sin^2\theta.
\end{eqnarray}
Similar constraints apply for other products. From the above
equation, one expects that on average
\begin{equation}
(\vec X_N'\cdot\vec X_i)^2 \sim \frac{1}{N-1}|\vec X_N'|^2\sin^2\theta
\end{equation}
which is suppressed by $1/(N-1)$. It is obvious that this
symmetric configuration is favored in cosmology. If one wants the
moduli to decay before BBN, then the most dangerous modulus is the
one with the smallest total decay width, which is bounded by the
average width. This gives rise to a strong constraint on the
geometry of the $G_2$ manifold, since the width is suppressed by
the number of moduli $N$. In the following discussion, we will
focus on this symmetric configuration.

In order to evaluate the decay width and the branching ratio, one
needs to know the typical values of these weighted dot products of
$\vec X_{N}$ and $\vec X_{i}$. To do the estimation, we generate a
set of fundamental parameters $a_i$, $N_i$, $N_i^{sm}$, $\xi_i$
and $\zeta_i$ randomly with the following conditions:
\begin{equation}
\sum_{i=1}^{N} a_i=\frac{7}{3}, \quad 1<N_i^{sm}<2, \quad 2<N_i<6, \quad -1<\xi_i<0, \quad -1<\zeta_i<0.
\end{equation}
The above ranges are chosen based on constraints arising from the
$G_2$ framework and our current understanding of the K\"{a}hler
metric of visible matter fields in realistic constructions. We
also impose the supergravity condition $V_7>1$ and volume of
three-cycle $V_Q^{sm}\approx 26$. The ranges of $N_i$ and
$N_i^{sm}$ are chosen such that the efficiency of the parameter
generation is maximized when the above constraints are imposed.
Due to our primitive understanding about the k\"{a}hler metric,
the modular weights (corresponding to $\xi_i$ and $\zeta_i$) are
taken randomly in the allowed range. We plot the distribution for
${\cal B}^{-2}(\vec X_N'\cdot \vec X_i)^2$ and $(\vec X_N'''\cdot
\vec X_i)^2$ in Fig.\ref{Fig:dist}, where we can see the typical
values are $2\times 10^{-4}$ and $20$. This result can be
understood from the very rough estimate $(\vec X_N'\cdot \vec
X_i)\sim \sqrt{a_i}\sim 1/\sqrt{N}$ and $(\vec X_N'''\cdot \vec
X_i)\sim 1/\sqrt{a_i}\sim \sqrt{N}$. The distribution of $((\vec
X_N'')^{\alpha}\cdot \vec X_i)^2$ is expected to be about the same
as $(\vec X_N'''\cdot \vec X_i)^2$, since they all have the same
structure. However, one should be aware that all the weighted dot
products are independent and so are not necessarily equal.
\begin{figure}[!h]
\begin{center}
   \includegraphics[width=7cm]{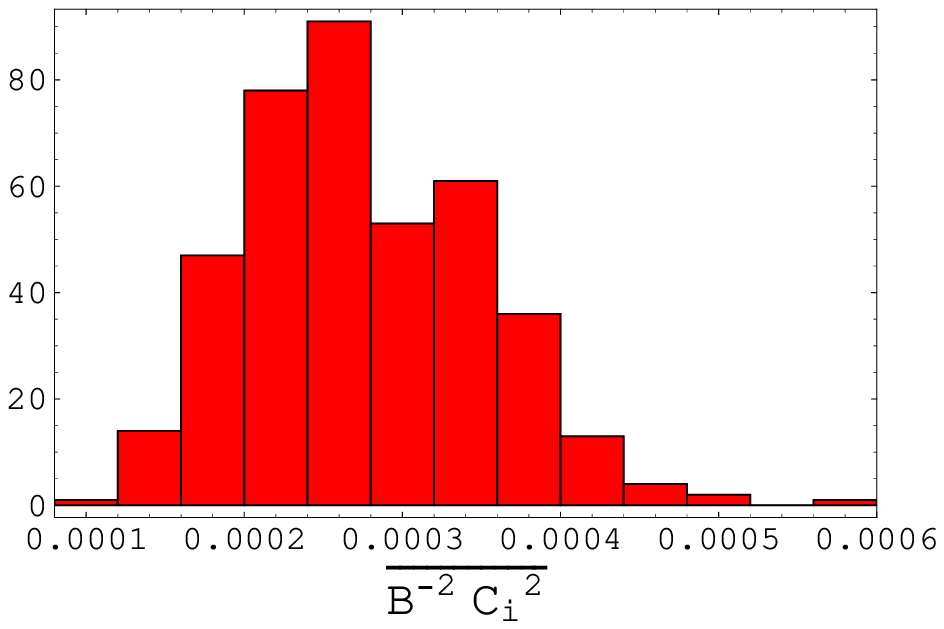}
   \includegraphics[width=7cm]{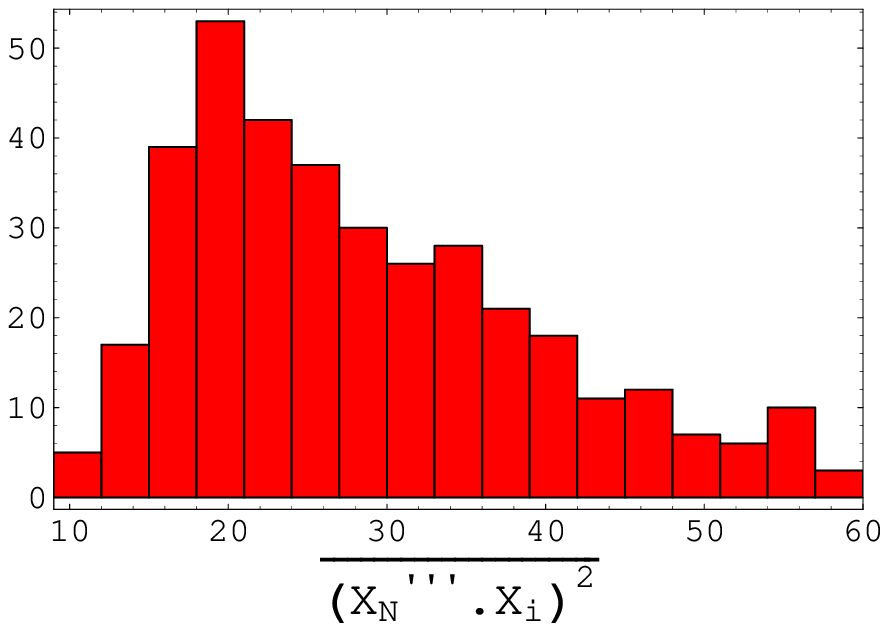}
\caption{Left: distribution of the average of ${\cal B}^{-2}(\vec X_N'\cdot \vec X_i)^2$. Right:
distribution of the average of the weighted dot product $(\vec X_N'''\cdot \vec X_i)^2$. \label{Fig:dist}}
\end{center}
\end{figure}

Now let us estimate the decay width for the light moduli. Consider
the first benchmark model of $G_2$-MSSM \cite{Acharya:2008zi} for
example, assuming the weighed dot products take their average
value, we find ${\cal A}_{1}^{X_i}\approx {\cal A}_{2}^{X_i}\sim
10^{-4}$, ${\cal A}_{3}^{X_i}\sim 7.3$ and ${\cal A}_{4}^{X_i}\sim
20.5$. To summarize, the main channels of interest for light
moduli decays and their partial widths are
$\Gamma(gg)=\Gamma(\tilde g\tilde g)\approx 0.024 \;{\rm
sec}^{-1}$, $\Gamma(\tilde t_R\tilde t_R)\approx 60 \;{\rm
sec}^{-1}$, $\Gamma(\tilde t_L\tilde t_L)= \Gamma(\tilde b_L\tilde
b_L)\approx 43 \;{\rm sec}^{-1}$ and $\Gamma(hh)\approx 412 \;{\rm
sec}^{-1}$. The total width is the sum of these partial width.
LSPs arise mainly from gauginos (including LSPs), $\tilde t\tilde
t$ and $\tilde b\tilde b$, so the LSP branching ratio is the sum
of the gaugino and squark channels divided by the total width. One
can see that the decay to higgs and scalar dominate the decay of
the light moduli. The total decay width is about $558\;{\rm
sec}^{-1}$ or the corresponding ${\cal D}_{X_i}=0.86$. The
branching ratio of the light moduli to LSP is about $26\%$. These
results should still be roughly correct for other benchmarks,
differing at most by $\mathcal{O}(1)$ since the dependence on the
mass spectrum is mild as seen from the explicit result of ${\cal
A}_i^{X_i}$. The main uncertainty arises from the deviation of
those weighted dot products from their typical values. To explore
the more general case, one can relax the condition that all the
weighted dot products are equal. Instead we choose:
\begin{eqnarray}
&&\vec X_N'''\cdot \vec X_i=(\vec X_N'')^{H_u}\cdot \vec X_i =(\vec X_N'')^{H_d}\cdot \vec X_i= \Pi_1 \\
&&(\vec X_N'')^{\tilde Q_3}\cdot \vec X_i=(\vec X_N'')^{\tilde
u_3}\cdot \vec X_i = \Pi_2
\end{eqnarray}
Then we vary $\Pi_1$ and $\Pi_2$ according to the distributions of
the weighted dot products in Fig.{\ref{Fig:dist}}. The
distribution for $D_{X_i}$ and the branching ratio to LSPs is
shown in Fig.{\ref{Fig:distWB}}. One can see that the branching
ratio has a very small variation, but the distribution of
$D_{X_i}$ has a long tail. In the paper, we will use $0.4 <
D_{X_i} < 4$ for concreteness, although other values may be
possible.

\begin{figure}[!h]
\begin{center}
   \includegraphics[width=7cm]{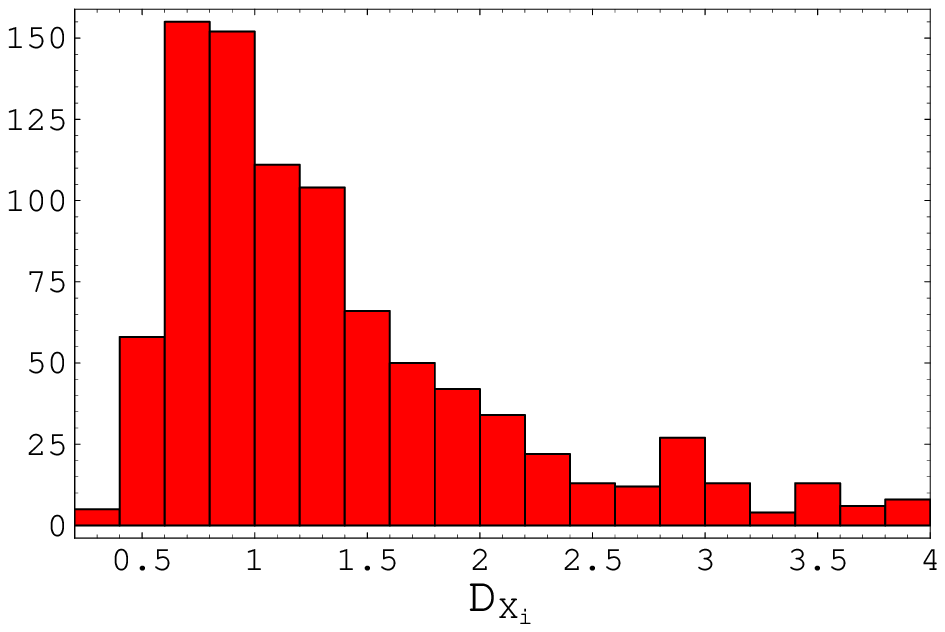}
   \includegraphics[width=7cm]{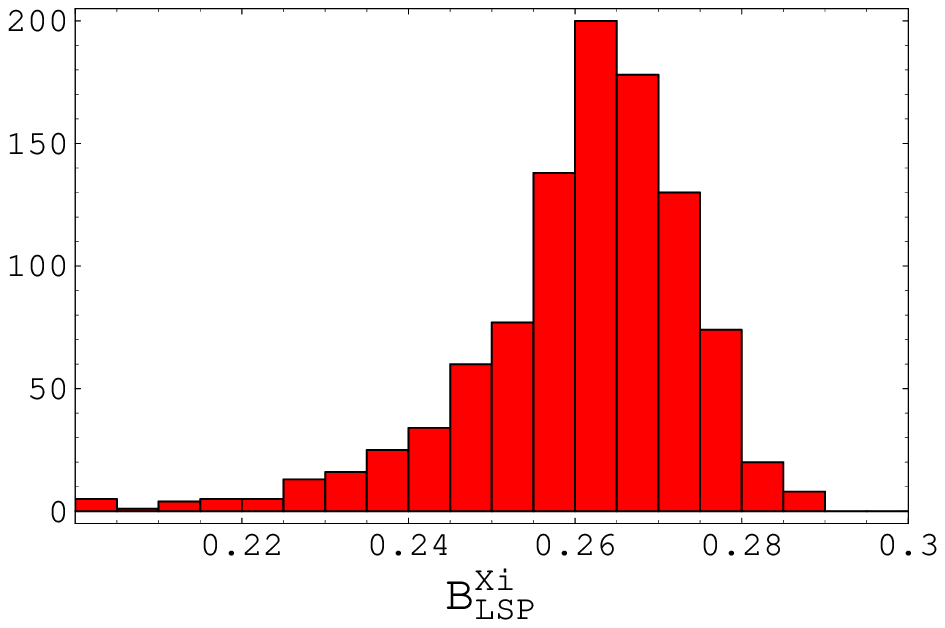}
\caption{Left: distribution of $D_{X_i}$. Right: distribution of
moduli branching ratio to LSP. \label{Fig:distWB}}
\end{center}
\end{figure}

For the heavy modulus $X_N$, the total decay width is
\begin{equation}
\Gamma(X_N)= \frac{7}{72\pi}\left(N_G{\cal A}_1^{X_N}+N_G{\cal
A}_2^{X_N}+{\cal A}_4^{X_N}\right)\frac{m_{X_N}^3}{m_p^2},
\end{equation}
where ${\cal A}_i^{X_N}$ corresponds to the decay to gauge bosons $gg$, gauginos $\tilde g\tilde g$ and higgs bosons, and are given by
\begin{eqnarray}
{\cal A}_1^{X_N}&=&\frac{9}{98}\\
{\cal A}_2^{X_N}&=&\frac{2}{9}\left(\frac{m_{3/2}}{m_{X_N}}\right)^2\left(\nu^2 b_1 b_2\right)^2\\
{\cal A}_4^{X_N}&=&Z_{\rm eff}^2\left(\vec X_N'''\cdot \vec
X_i\right)^2.
\end{eqnarray}
In the above result, we have not included the contributions from
the decay to non-higgs scalars and fermions since they are
suppressed by $(m_{3/2}/m_{X_N})^4$ and $(m_f/m_{X_N})^2$ given
the large mass of the heavy modulus $m_{X_N}\sim 600\times
m_{3/2}$. Taking benchmark 1 of $G_2$-MSSM in
\cite{Acharya:2008zi} and typical values for weighted dot
products, we get ${\cal A}_1^{X_N}\approx 0.1$, ${\cal
A}_2^{X_N}\approx 0.01$ and ${\cal A}_4^{X_N}\approx 50$. The
total width is about $3\times 10^{10}\;{\rm sec}^{-1}$ or the
corresponding $D_{X_N}\approx 1.6$. The branching ratio to LSPs is
about $3\times 10^{-3}$.

The decays of moduli to gravitinos is also very important. The
decay of a modulus to gravitinos can be calculated using the
following formula: \cite{Kawasaki:2006hm}
\begin{eqnarray}
  \Gamma(X\rightarrow 2\psi_{3/2})\simeq \frac{|{\cal
  G}_{X}^{(eff)}|^2}{288\pi}\frac{m_{X}^5}{m_{3/2}^2
  m_p^2}\label{Eq:gravitino}
\end{eqnarray}
where ${\cal G}_{X}^{(eff)}$ is the effective coupling of the
modulus field to gravitinos which includes effects of moduli
mixing. For the heavy modulus, the coupling arises from the mixing
with meson field, since the goldstino is mainly the fermionic
partner of the meson. Since the heavy modulus is much heavier than
the meson a rough estimate\footnote{There could be an additional suppression in special cases as discussed in \cite{hep-ph/0605091,Dine:2006ii}.  
We thank Fuminobu Takahashi for discussions regarding this point.} gives
${\cal G}^{(eff)}_{X}\sim m_{3/2}/m_{X_N}$.
Therefore, for the heavy modulus, the decay rate to gravitino is
\begin{eqnarray}
   \Gamma(X_N\rightarrow 2\psi_{3/2})\sim \frac{1}{288\pi}\frac{m_{X_N}^3}{
  m_p^2}
\end{eqnarray}
This corresponds to $B_{3/2}^{X_N}\sim 7\times 10^{-4}$. In
addition, since the heavy modulus decays much earlier than other
moduli, the gravitino produced will be diluted by the subsequent
moduli decays. So this estimate is enough for our discussion of
gravitino problem. For both the light moduli and the meson fields,
the decay to gravitino is kinematically suppressed since $m_{X_i},
m_{\phi_0} \approx 2m_{3/2}$.

\subsection{Decay Width of the Meson}

The total decay width of meson modulus is:
\begin{equation}
\Gamma(\delta\phi_0)\equiv
\frac{D_{\phi}m_{\phi}^3}{m_p^2}=\frac{1}{72\pi}\left(N_G{\cal
A}_1^{\phi_0}+{\cal A}_2^{\phi_0}+{\cal
A}_3^{\phi_0}\right)\frac{m_{\phi}^3}{m_p^2},
\end{equation}
where ${\cal A}_i^{\phi}$ corresponds to the decay to gauginos
$\tilde g\tilde g$, non-higgs scalar $\tilde f\tilde f$ and light
higgs bosons $hh$, and are given by:
\begin{eqnarray}
{\cal A}_1^{\phi_0}&=&\frac{1}{2\phi_0^2}{\cal F}^2 \left(\frac{m_{3/2}}{m_{\phi}}\right)^2,\\
{\cal A}_2^{\phi_0}&=&\sum_{\alpha} 27\phi_0^2 K_3^2 Z_{\rm eff}^4\left((1+Z_{\rm eff}^{-2})(1+\frac{2}{3\phi_0^2})+\frac{2\cal F}{3\phi_0^2}\sum_{i=1}^{N}\zeta_i\right)^2\nonumber\\
&\times&\left(\frac{m_{3/2}^4}{m_{\phi}^4}\right)
\left(1-4\frac{m_{\tilde f_{\alpha}}^2}{m_{\phi}^2}\right)^{1/2},\\
{\cal A}_3^{\phi_0}&=&\frac{9}{2}\phi_0^2\left(\frac{m_{3/2}^4}{m_{\phi_0}^4}\right)\bigg[Z_{\rm eff}^2\left((1+Z_{\rm eff}^{-2})(1+\frac{2}{3\phi_0^2})
+\frac{2\cal F}{3\phi_0^2}\sum_{i=1}^{N}\zeta_i\right)(\sin^2\alpha+K_1\cos^2\alpha)\nonumber\\
&-&K_2Z_{\rm eff}\left(2(1+\frac{2}{3\phi_0^2})+\frac{\cal F}{3\phi_0^2}\sum_{i=1}^{N} (\xi_i^{H_u}+\xi_i^{H_d})\right)\sin2\alpha\bigg]^2.
\end{eqnarray}
Here, as discussed in the last subsection, the low scale couplings
to third-generation squarks are dominantly generated from RG
running and are related to the coupling $g_{\phi H_u H_u}$ by a
factor $K_3\sim 0.1$. For the first benchmark of $G_2$-MSSM in
\cite{Acharya:2008zi} and taking the simplest assumption
$\xi_i=\zeta_i=-1/2$, we get ${\cal A}_1^{\phi_0}\approx 13.7$,
${\cal A}_2^{\phi_0}\approx 2.9\times 10^4$ and ${\cal
A}_3^{\phi_0}\approx 1.3\times 10^5$. One can see that this result
is enhanced from the naive estimate by the total number of moduli
$\sim N$ and large (hidden-sector) three-cycle volume $\nu$. The
total decay width is about $4.6\times 10^5\; {\rm sec}^{-1}$,
corresponding to $D_{\phi}=711$. The branching ratio to LSPs is
about $18\%$. Again, this can change by $\mathcal{O}(1)$ for other
benchmarks.

\end{document}